\documentclass[letterpaper]{article} 
\usepackage{aaai25}  
\usepackage{times}  
\usepackage{helvet}  
\usepackage{courier}  
\usepackage[hyphens]{url}  
\usepackage{graphicx} 
\usepackage{natbib}  
\usepackage{caption} 
\usepackage{algorithm}
\usepackage{algorithmic}

\usepackage{newfloat}
\usepackage{listings}
\DeclareCaptionStyle{ruled}{labelfont=normalfont,labelsep=colon,strut=off} 
\floatstyle{ruled}
\newfloat{listing}{tb}{lst}{}
\floatname{listing}{Listing}
\usepackage{multirow}
\usepackage{multicol}
\usepackage{booktabs}
\usepackage{url}
\usepackage{amsmath}
\usepackage{xcolor}
\usepackage{amssymb}

\usepackage{xspace}
\newcommand{\toolW}{{AdaPatcher}}
\newcommand{\tool}{{$\boldsymbol{AdaPatcher}$}\xspace}

\newcounter{stage}
\newcounter{stage2}
\title{Less is More: Adaptive Program Repair \\with Bug Localization and Preference Learning}
\author{\bf ~~Zhenlong Dai$^{1}$, 
        ~~Bingrui Chen$^{2}$,
        ~~Zhuoluo Zhao$^{3}$, \\
        ~~Xiu Tang$^{1}$, 
        ~~Sai Wu$^{1}$, 
        ~~Chang Yao$^{1}$, 
        ~~Zhipeng Gao$^{1}$\thanks{Co-corresponding authors.},  
        ~~Jingyuan Chen$^{1*}$\\
        \textsuperscript{1}Zhejiang University,
        \textsuperscript{2}Hohai University,
        \textsuperscript{3}Guizhou University\\
        {\tt\small \{zhenlongdai,tangxiu,wusai,changy,zhipeng.gao,jingyuanchen\}@zju.edu.cn}
        \tt\small ChenBingrui@hhu.edu.cn,\tt\small ie.zlzhao21@gzu.edu.cn
        }
\usepackage{bibentry}

\begin{document}

\maketitle

\begin{abstract}
Automated Program Repair (APR) is a task to automatically generate patches for the buggy code. 
However, most research focuses on generating correct patches while ignoring the consistency between the fixed code and the original buggy code. 
How to conduct adaptive bug fixing and generate patches with minimal modifications have seldom been investigated. 
To bridge this gap, we first introduce a novel task, namely \textbf{AdaPR} (\textbf{\underline{Ada}}ptive \textbf{\underline{P}}rogram \textbf{\underline{R}}epair). 
We then propose a two-stage approach \tool (\textbf{\underline{Ada}}ptive \textbf{\underline{Patch}} G\textbf{\underline{e}}nerato\textbf{\underline{r}}) to enhance program repair while maintaining the consistency. 
In the first stage, we utilize a Bug Locator with self-debug learning to accurately pinpoint bug locations. 
In the second stage, we train a Program Modifier to ensure consistency between the post-modified fixed code and the pre-modified buggy code. 
The Program Modifier is enhanced with a location-aware repair learning strategy to generate patches based on identified buggy lines, a hybrid training strategy for selective reference and an adaptive preference learning to prioritize fewer changes. 
The experimental results show that our approach outperforms a set of baselines by a large margin, validating the effectiveness of our two-stage framework for the newly proposed AdaPR task. The code and dataset are available at \url{https://github.com/zhenlongDai/AdaPatcher}.
\end{abstract}

%
\begin{figure}[t]
\includegraphics[width=0.48\textwidth]{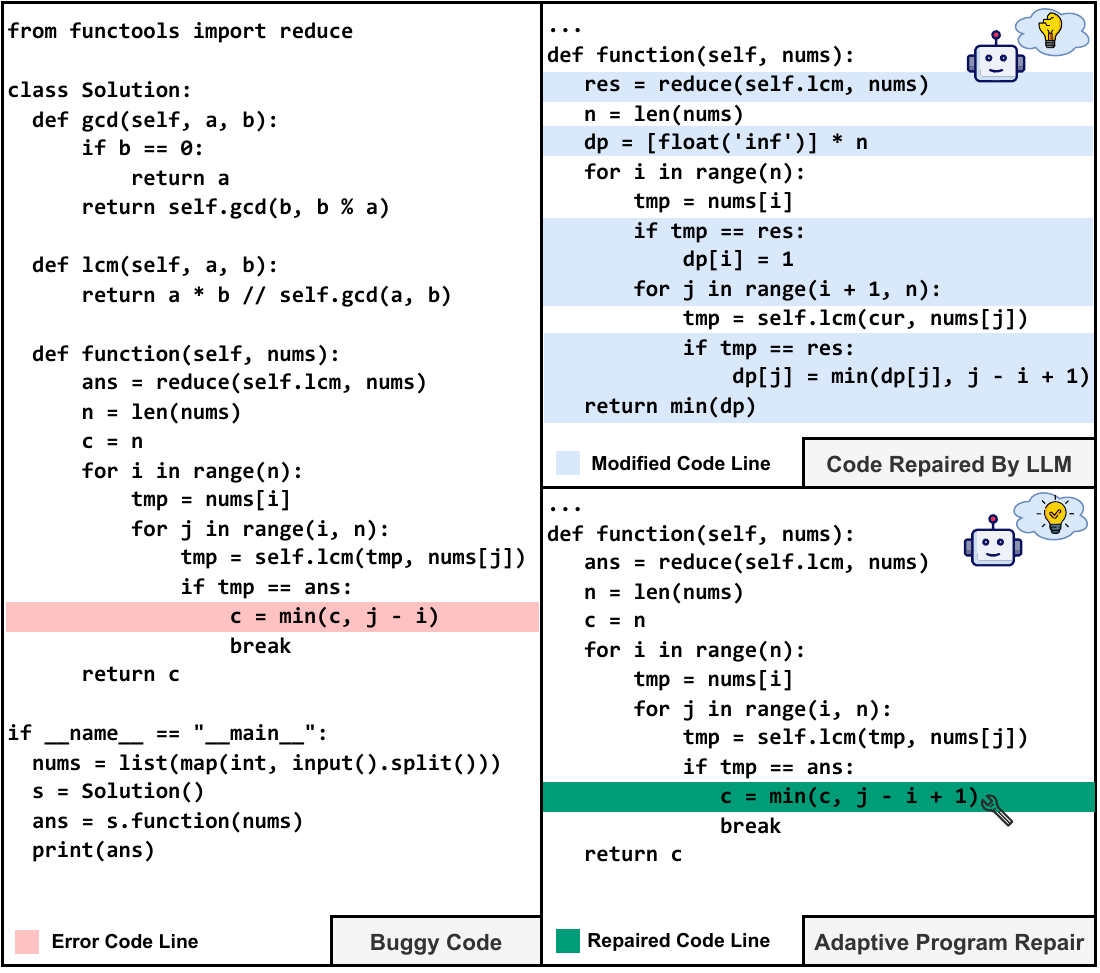}
\vspace{-1em}
\caption{Example of AdaPR. The adaptive repaired code is correct and minimizes code modifications.
\vspace{-1em}
}
\label{fig:introduction}
\end{figure}
\section{\textbf{\textsc{Introduction}}}
As software systems become more and more prevalent in everyday life, software bugs also become inevitable. 
These software bugs can potentially cause security issues or even financial losses~\cite{shahriar2012mitigating,dissanayake2022software,krasner2021cost}. 
Usually, developers need to fix these buggy codes manually by spending a significant amount of time and effort. 
To alleviate developers' burden for bug fixing, Automated Program Repair (APR) has been introduced to automatically generate patches given the original buggy code. 
APR techniques take a buggy code and a correct specification as input, aiming to generate the fixed program satisfying the given specifications.

Nowadays, inspired by the promising performance of Large Language Models (LLMs) in code generation~\cite{jiang2024survey,li2023starcoder} and code understanding~\cite{li2022competition,chen2021evaluating}, researchers have applied LLMs to perform the APR task~\cite{ye2022selfapr,fan2023automated,jin2023inferfix} and demonstrated remarkable results. 
However, most studies focus on generating \textit{correct} patches, the \textit{consistency} between the fixed code and the buggy code is often ignored and cannot be guaranteed, which greatly hinders the practical use of LLM for bug fixing. 
Consider the practical scenario in Fig.~\ref{fig:introduction} as an example, Alice is a developer, she implemented the \texttt{Solution} class to achieve her goal. 
Nonetheless, her program failed to pass the tests which indicates potential bugs within her written code. 
The error message suggested ``AssertionError: Expected output is 4, but the received output is 3''. 
Alice tried to use LLM to help her fix this bug by feeding LLMs with the original buggy code and error message.
However, the patch generated by LLMs overwrote most code lines in \texttt{function} (\textit{e.g.}, colored in blue). 
It is difficult for Alice to accept this patch because the generated code is too far from her original written one. 
The extensive modifications made by LLMs make the fixed code hard to trace and understand. 
As a result, Alice refused to integrate this patch into her codebase. 


To address this gap, we propose a new task in this paper, namely \textbf{Adaptive Program Repair}, denoted as \textbf{AdaPR}. 
Different from APR, AdaPR not only aims to generate ``\textit{correct}'' patches for the buggy code, but also aims to generate ``\textit{consistent}'' patches with minimal modifications. 
More formally, given the buggy code and the correct specifications (\textit{e.g.}, failed test cases), AdaPR adaptively fixes the buggy program with the least possible changes to satisfy the given specification. 
For example, in Fig.~\ref{fig:introduction}, Alice can fix this bug by only changing one line of code, \textit{i.e.}, from \texttt{c = min(c, j - i)} to \texttt{c = min(c, j - i + 1)}. 
This newly generated patch aligns with her design intentions and existing code structures, the consistency between the fixed code and the original buggy code makes Alice easy to understand the code changes and increases her confidence of this fix pattern. 
Consequently, Alice accepted this patch without a doubt and incorporated it into her codebase.


So far, the existing studies focus on generating \textit{correct} patches, there is no research investigating how to adaptively fix buggy code with minimal modifications. 
AdaPR is a non-trivial task regarding the following key challenges: (i) \textbf{\textit{Where to fix:}} Identifying the precise location(s) where the bug has been introduced is challenging. 
When a bug occurs, different parts of the program may exhibit abnormal behaviors according to the bug. 
To fix the bug adaptively, AdaPR first requires locating the root cause of the problem and pinpointing the exact buggy line(s) that need modifications. 
(ii) \textbf{\textit{How to fix:}} Generating patches with minimal modifications is challenging. 
Because LLMs are typically trained on general programming corpus (\textit{e.g.}, comment-code pairs, question-solution pairs), LLMs' primary goal is to generate correct and functional code. 
Regarding program repair, LLMs tend to repair a program by rewriting it from scratch without considering the existing code structure or semantics. 
AdaPR requires the patches to be both correct and consistent. 
In other words, the generated patches should involve as few modifications as possible while still addressing bugs effectively. 
How to fix the program incrementally and adaptively is another challenge in this work.

To tackle the above challenges, we propose a novel two-stage approach named \tool, which is designed to patch a buggy program correctly and consistently.  
To address the first \textit{where to fix} challenge, we propose a diff-based component, namely Bug Locator, to pinpoint the exact bug locations within the buggy code. 
Specifically, for a passed program and a failed program written by the same developer, we first record the run-time values of different variables respectively. 
Following that, we teach LLM to do self-debug learning to identify bug locations, i.e., the LLM is guided to debug and analyze the differences (\textit{e.g.}, code deletions, modifications) between the passed program and the failed program and finally determine which code line causes the test failures. 
To address the second \textit{how to fix} challenge, we design a Program Modifier component for our second stage. 
The Program Modifier is enhanced with three techniques to ensure the consistency and correctness of the fixed program and the original buggy program. 
Particularly, to avoid LLMs repairing the program from scratch, we leverage location-aware repair learning to generate patches based on the identified buggy lines. 
To reduce the negative effects of the incorrect bug locations, we propose a hybrid training strategy that enables the Program Modifier to selectively reference bug locations instead of blindly modifying them. 
Moreover, to make code changes as small as possible, the Program Modifier is trained to generate fewer modifications by adaptive preference learning. 

In summary, our paper makes the following contributions: 
(1) We first propose a novel task, namely AdaPR, to fix buggy code with minimal modifications. 
This newly proposed task aims to produce both correct and consistent code patches, which is more practical in real-world software development; 
(2) We build a dataset with over 50K $\langle buggy~code, test~case, fixed~code \rangle$ data samples, to the best of our knowledge, it is the first large dataset for this task; and
(3) We present a novel model, named \tool, to perform the AdaPR task. 
\tool is based on LLMs and introduces several customized improvements to effectively handle \textit{where to fix} and \textit{how to fix} challenges. 
The experimental results show the effectiveness of our model over a set of baselines, showing its potential to enhance automated program repair while reducing modifications at the same time. 
We hope our study can lay the foundations for this research topic.

\section{\textbf{\textsc{Related Work}}}

Recent advancements in LLMs have spurred their integration into automated program repair~\cite{sobania2023analysis,xia2023automated,jiang2023impact,paul2023enhancing}. Enhancing code LLMs with feedback mechanisms has demonstrated potential~\cite{miceli2023larger}, particularly through feedback from tools like compilers ~\cite{bouzenia2023tracefixer,xia2022less} such as traces or test diagnostics. CoT reasoning loop~\cite{yao2022react} has been used to predict repair actions based on interactive feedback from debuggers. NExT~\cite{ni2024next} focuses on tuning LLMs to reason with pre-existing execution information. Additionally, LLMs can generate natural language explanations for errors~\cite{chen2023teaching,zhang2022coditt5}, offering another valuable form of feedback. Self-improvement methods iteratively refine code generated by LLMs using CoT reasoning over self-provided feedback~\cite{madaan2024self,zhang2023self}. 
CoFFEE~\cite{moon2023coffee} uses LLMs to generate natural language explanations for errors.
Existing approaches in automated program repair focus on accuracy, while our research aims to enhance program repair with minimal modifications.

\section{\textbf{\textsc{Methodology}}}

In this section, we introduce a novel two-stage framework \tool, aimed at enhancing program repair while maintaining the consistency. The first stage employs a Bug Locator to identify the root cause of bugs and pinpoint the buggy code lines in the form of \texttt{Code Diff}\footnote{\texttt{Code Diff} refers to the differences between the buggy and correct code.}. The second stage utilizes a Program Modifier to adaptively propose fixes for the identified buggy code lines. This approach prioritizes patches that require minimal changes, thus preserving the cleanliness and maintainability of the codebase.



\begin{figure*}[thb!]
\centering
\includegraphics[width=0.95\textwidth]{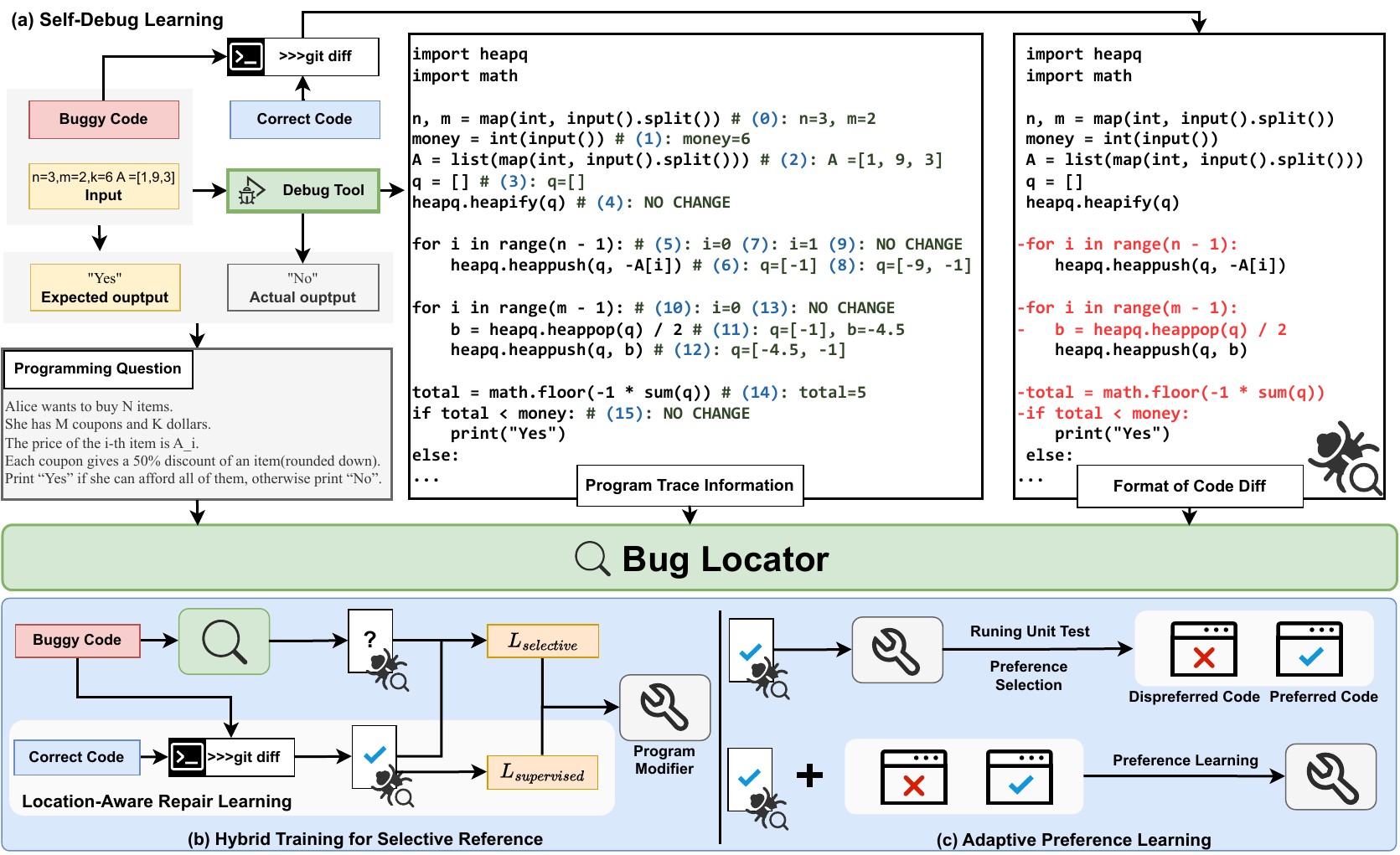}
\vspace{-0.5em}
\caption{Overview of \tool. (a) Illustration of the Self-Debug Learning process. (b) Illustration of the Hybrid Training for Selective Reference process. (c) Illustration of the Adaptive Preference Learning process. 
} 
\label{fig:model_first}
\vspace{-1em}
\end{figure*}

\subsection{Task Definition}

Given a specific programming task $q$, a buggy code $c$, and a correct specification $s$, the objective is to generate a fixed version of the buggy code, denoted as $y$, which satisfies the specification $s$. The fixed version $y$ should maintain consistency with the surrounding code and require minimal modifications to the original code $c$. 

\subsection{Stage \Roman{stage}: Bug Locator}
LLMs demonstrate strong code comprehension capabilities~\cite{nam2024using}; however, they often struggle with accurately identifying and describing code bugs~\cite{olausson2023self}, particularly in pinpointing buggy lines. 
To address this challenge, we propose a diff-based approach that simplifies and clarifies bug locations for LLMs. 
As shown in Fig.~\ref{fig:model_first}, bug locations are aligned with the corresponding buggy lines both semantically and structurally, with a `-' symbol prefix indicating the need for deletion or correction. 
The form of \texttt{Code Diff}  simplifies and clearly pinpoints bug locations by providing a structured and explicit indication of where changes are needed. 
This makes it easier for the Bug Locator to focus their attention on the relevant buggy portions of the code.

Additionally, since LLMs often lack an understanding of program execution at runtime, identifying and locating runtime bugs is challenging.
To address this, we propose a novel self-debug learning to enhance the Bug Locator's ability to identify and locate runtime errors.

\subsubsection{Format of \texttt{Code Diff}.}
Given a buggy code 
$c = \{c_1,c_2,...,c_n\}$ and a corrected code $y$, a diff file is generated using  Git\footnote{\url{https://git-scm.com/}} by comparing $c$ and $y$. Lines marked with a `-' symbol in the diff file are identified as buggy lines $L$. The diff file $d =\{d_1,d_2,...,d_n\}$ is created by prefixing buggy lines in $c$ with a `-' symbol:
\begin{align}
d_i = \begin{cases}
\text{\textless space\textgreater} \cdot c_i, & c_i \notin L \\
\phantom{spa} \text{`-'} \ \ \ \ \ \cdot c_i, & c_i \in L
\end{cases}
\end{align}
where $d_i$ is the $i$-th line of the diff file, and both $d$ and $c$ contain $n$ lines. The symbol $\cdot$ represents string concatenation, and $\text{\textless space\textgreater}$ denotes a whitespace character.


\subsubsection{Self-Debug Learning.} 
Certain bugs manifest only during runtime, necessitating an understanding of program execution. LLMs often struggle with these bugs due to their training on the static textual form of code. Drawing inspiration from the practice of rubber duck debugging~\cite{parreira2023robot, ni2024next}, we introduce self-debug learning to enhance the Bug Locator $\theta$'s ability to identify and localize runtime bugs.

Specifically, given a buggy code $c$ and a corresponding failed test case $t$ from the correct specification $s$, the code is executed with $t$ to capture the actual output. The program's I/O data, denoted as $D_t$, includes both input and expected/actual output. Additionally, using the Python `\texttt{traceback}' module\footnote{\url{https://docs.python.org/3/library/traceback.html}}, as shown in Fig.~\ref{fig:model_first}, we capture the variable states at each executed line and record the execution order (\textit{e.g.}, colored in blue) to create program trace information $R_t$.

To facilitate LLM comprehension of program trace information, $R_t$ is formatted as compact inline code comments (\textit{e.g.}, colored in green) that do not disrupt the code structure: 
1) Comments display only variables that change after each line's execution, marking each execution step; and 2) Loop trace information is compressed using ellipses for large iteration counts. 
As shown in Fig.~\ref{fig:model_first}, $R_t$ is structurally aligned with diff-based file $d$, providing a coherent format for the Bug Locator to identify and localize errors. 
The self-debug prompt is then constructed as: 
\begin{figure}[!thb]
\centering
\fcolorbox{black}{gray!6}{%
\parbox{0.98\linewidth}{%
\small
\noindent$\bullet$ \textbf{Instruction}: Given a programming question and a corresponding piece of buggy code written in \textless language\textgreater, please provide a program repair proposal for the buggy code.  Use `-' to represent the line that may need to be deleted or modified.\\
\rule{\linewidth}{0.48pt} 
\vspace{-0.2em} 
$\bullet$ \textbf{Programming Task}: $q$\\
$\bullet$ \textbf{Buggy Code}: $c$\\
$\bullet$ \textbf{Execution Information of Failed Test Case}:\\
\vspace{-0.1em} 
\hspace{1em}\textbf{I/O data}: $D_t$\\
\vspace{-0.1em}
\hspace{1em}\textbf{Program Trace Information}: $R_t$
}%
}

\label{fig:Fault-Driven-prompt}
\vspace{-0.5em} 
\end{figure}

The objective of self-debug learning is to minimize the negative log-likelihood of the \texttt{Code Diff} file $d$ by utilizing the prompt:
\begin{align}
\mathcal{L}_{\text{BL}}= - \!\!\!\!\!\!\!\sum_{(q,c,t,d) \sim \mathcal{D}}\!\!\!\!\! \log P_\theta(d|q,c,D_t,R_t),
\end{align}
where all repair instances $(q,c,t,d,y)$ form the dataset $\mathcal{D}$, and $P_\theta$ represents the probability distribution over the LLM's vocabulary.

\subsection{Stage \Roman{stage2}: Program Modifier}
Repairing code with few modifications requires understanding the modification process of buggy code. However, LLMs typically struggle with this process since they may not have the knowledge to make informed decisions about which changes to make in order to repair the code effectively. 
To address this challenge, we introduce location-aware repair learning, which directs the Program Modifier to focus on buggy areas identified in the first stage. 
Recognizing the possibility of incorrect bug locations produced by the Bug Locator, we propose a hybrid training strategy to prevent the Program Modifier from making unnecessary modifications, thereby improving repair accuracy. 
Additionally, to further reduce the extent of modifications, we train the Program Modifier to align with the preference for fewer changes through adaptive preference learning.
\subsubsection{Location-Aware Repair Learning.}
To explicitly capture the modification process, we propose Location-Aware Repair Learning, which trains the Program Modifier to make precise fixes by focusing on identified buggy areas.
Given the bug locations and correct code, we guide the Program Modifier to make corrections without altering unrelated code.

Specifically, the Program Modifier $\phi$ is trained using supervised learning to predict the correct code $y$ as follows:
\begin{align}
\mathcal{L}_{\text{supervised}}= - \!\!\!\!\!\!\!\sum_{(q,c,d,y) \sim \mathcal{D}}\!\!\!\!\! \log P_\phi(y|q,c,d),
\end{align}
 where $P_\phi \in \mathbb{R}^{|\mathcal{V}|}$ is the probability distribution on the LLM's vocabulary.

\subsubsection{Hybrid Training for Selective Reference.}
The Bug Locator may generate incorrect bug locations, potentially leading the Program Modifier to fail in fixing the bugs.
To address this issue, we propose a hybrid training strategy for selective reference, which further trains the Program Modifier to repair code based on bug locations that may be incorrect.
The training strategy enhances the Program Modifier's selective reference to bug locations provided by the Bug Locator instead of blindly modifying them.

Specifically, 
the dataset $\mathcal{D}$ is split into $\mathcal{D}_1$ and $\mathcal{D}_2$, 
with the data volumes satisfying $|\mathcal{D}_1| : |\mathcal{D}_2| = 1:k$, where $k$ is ratio parameter.
Each instance $(q,c,d,y) \in \mathcal{D}_2$ is processed by the Bug Locator $\theta$ to generate new labels $\hat{d}$:
\begin{align}
\hat{d} = \text{LLM}_\theta(q,c,D_t,R_t).
\end{align}
Then we construct a new dataset $\mathcal{D}_2'$:
\begin{align}
\mathcal{D}_2' = \{(q, c,d,\hat{d},y) \mid (q,c,d,y) \in \mathcal{D}_2\}.
\end{align}
The loss function of selective reference for the Program Modifier is:
\begin{align}
\mathcal{L}_{\text{selective}}= - \!\!\!\!\!\!\!\!\!\!\!\!\sum_{(q,c,d,\hat{d},y) \sim \mathcal{D}_2'}\!\!\!\!\!\!\!\!\!\!(\log P_\phi(y|q,c,d)+\log P_\phi(y|q,c,\hat{d})).
\end{align}
We jointly train the Program Modifier using supervised learning data and selective learning data to maintain its repair capability and enhance its selective reference ability:
\begin{align}
\mathcal{L}_{\text{Hybrid}}=  \mathcal{L}_{\text{supervised}}(\mathcal{D}_1)  +
\mathcal{L}_{\text{selective}}(\mathcal{D}_2'),
\end{align}
where $\mathcal{L}(\cdot)$ denotes the loss function applied to the respective dataset during training.

\subsubsection{Adaptive Preference Learning.}
Even when fixing the same bug, different methods can result in varying extents of code changes. 
To prioritize fewer modifications during the repair process, we draw inspiration from Direct Preference Optimization (DPO)~\cite{rafailov2024direct}, which steers LLMs to match specific preferences. 
Building on DPO, we propose an adaptive preference learning mechanism that guides LLMs to reduce the extent of code modifications further.
Given two codes, differing in the extent of modifications, we utilize preference learning to guide the Program Modifier in learning preference for fewer modifications, as shown in Fig.~\ref{fig:model_first}.

Specifically, we obtain preference pairs $(y^+,y^-)$, representing the preferred (\textit{i.e.}, the correct version with fewer modifications) and dispreferred (\textit{i.e.}, the incorrect version with more extensive modifications) codes generated by the Program Modifier $\phi$ after running the unit test. 
The preference set $\mathcal{D}_{p}$ consists of preference pairs $(q,c,d,y^+,y^-)$.
Based on the preference set $\mathcal{D}_{p}$, we apply DPO-Positive learning~\cite{pal2024smaug} to enhance the Program Modifier $\phi$, iterating on its training to derive $\phi^*$ that prioritize repairs requiring fewer modifications.
Formally, the training objective of Program Modifier $\phi^*$ is defined as:
\begin{equation}
\begin{split}
\mathcal{L}_{\text{repair}}(\phi^*; \phi)&=  - \mathbb{E}_{(x,y^+,y^-) \sim \mathcal{D}_{p}} \\
&\log \sigma [ r(x, y^+) - r(x, y^-) - g(x,y^+)],
\end{split}
\end{equation}
where $\sigma$ denotes the logistic function, $(q,c,d)$ is simplified as $x$, and $r$ is the reward function on the generated code implicitly defined by $\phi^*$ and $\phi$, with a hyperparameter $\beta$ to control the deviation from $\phi$ as:
\begin{equation}
r(x,y) = \beta \log \frac{P_{\phi^*}(y|x)}{P_{\phi}(y|x)}.
\end{equation}
And $g$ denotes the penalty term within the log-sigmoid to encourage maintaining a high log-likelihood of the preferred code:
\begin{equation}
g(x,y^+) = \lambda \cdot \max( 0, \log  \frac{P_\phi(y^+|x)}{P_{\phi^*}(y^+|x)} ),
\end{equation}
where $\lambda$ is a hyperparameter. After training, the Program Modifier is optimized to increase the probability of generating the preferred code, thereby achieving effective repairs with fewer changes.

\section{\textbf{\textsc{Experiments}}}
\subsection{Experimental Setups}
\subsubsection{Dataset.} 
We construct the first dataset, named \textbf{ACPR} (Accuracy-Consistency Program Repair) for our AdaPR task, which aims to evaluate the generated patches from accuracy (\textit{i.e.}, fixing bugs correctly) and consistency (\textit{i.e.}, minimizing modifications). 
Specifically, our dataset is collected from CodeNet~\cite{puri2021codenet}, which contains submissions of programming problems from different users. 
For a given buggy program, we pair it with a randomly selected failed test case (a test case includes a test input and an expected output) as well as a passed program from the same user's submission for the same programming problem, making a $\langle buggy~code, failed~test~case, passed~code \rangle$ triplet sample. 
The whole dataset contains 52,168 triplet data samples.
We then split our dataset into train/validation/test sets by the ratio of 8:1:1, ensuring that any particular programming problem appears in only one of them to avoid data leakage problems. 
To prevent overfitting code data from the same programming problem and to ensure fairness in evaluation, we balance the dataset by capping the maximum number of pairs per problem at 150/10/20 in the train/validation/test sets.
The overall statistics of the dataset are given in Table \ref{Table-1}. Further details can be found in the Appendix.

\begin{table}[thb!]
\small
\setlength{\tabcolsep}{1mm} %
\centering
\begin{tabular}{lcccccc}
\toprule 
\multirow{2}{*}{\textbf{Sets}} & \multirow{2}{*}{\textbf{triplet}} & \multirow{2}{*}{\textbf{problems}} & \multirow{2}{*}{\textbf{users}} & \multicolumn{3}{c}{\textbf{information per problem} } \\
& & & & \textbf{triplet}& \textbf{code lines}& \textbf{test cases}\\
\midrule 
train & 50023 & 688 & 13701   & 73 &28 &80\\
val & 927 & 110 & 783  & 8 &34 &87\\
test & 1668 & 110 & 1247  & 15 &40 &84\\
\bottomrule
\end{tabular}

\caption{Dataset Statistics.}
\label{Table-1}
\vspace{-1em}
\end{table}

\subsubsection{Evaluation Metrics.} 
To thoroughly evaluate a model's performance regarding our AdaPR task, we adopted the following evaluation metrics: 
(1) \textbf{Code Accuracy Rate (Acc)}: It represents the percentage of code that successfully passes all test cases of the programming problem~\cite{muennighoff2023octopack}. 
(2) \textbf{Code Improvement Rate (Improve)}: 
This metrical measures the average improvement rate for each piece of buggy code. It calculates the proportion of additional test cases passed after the buggy code is modified. 
The calculation equation for the improvement rate of the $i$-th fixed code follows:
\begin{equation}
I_i = \frac{\chi (\mathcal{A})\times n}{m},     
\end{equation}
where $\chi(\cdot)$ is an indicator function that returns 1 if the condition inside the parentheses is true, and 0 otherwise. 
$\mathcal{A}$ is true if all previously passing test cases still pass after the code modification, and false otherwise.
$n$ denotes the number of cases that additional pass after repair and $m$ represents the number of test cases that failed previously. 
The value of $i$-th fixed code is $I_i$ if the code passes all test cases passed by the buggy code, and 0 otherwise.
(3) \textbf{Failed Repair Rate (FR)}: It counts the proportion of the generated code that fails to pass the previously passed cases, which is calculated as follows:
\begin{equation}
    FR=\frac{\sum_{i=1}^{|D|}\chi (\mathcal{B}_i)}{|D|},
\end{equation}
where $|D|$ is the number of pieces of code,
$\chi(\cdot)$ is an indicator function. 
$\mathcal{B}_i$ is true if the $i$-th piece of code causes the previously passed cases to fail, and false otherwise. 
(4) \textbf{Code Consistency Rate (Consistency)}: It calculates the proportion of lines of code that are preserved after modification. 
It is defined as follows:
\begin{equation}
    Consistency=\frac{r}{k},
\end{equation}
where $k$ indicates the total number of code lines in the fixed code, and $r$ indicates the number of code lines preserved in the after-modification code. Further details can be found in the Appendix.

\begin{table*}[thb!]
\small
\setlength{\tabcolsep}{3mm}
\centering
\begin{tabular}{c|clccccc}
\hline 
Repair&Framework Structure&Methods&Bug Localization& Acc& Improve& Consistency &FR \\
\hline 
\multirow{6}{*}{CodeLlama} & \multirow{4}{*}{End-to-end} & Instruction&- &10.67&11.67&54.96&30.70\\
&&CoT&-&10.43&11.78&57.06&33.33\\
&&Few-shot Learning&-&9.35&10.50&40.20&31.18\\
&&Fine-Tuning&-&28.12&30.34&51.16&14.21\\
\cline{2-8} 
&\multirow{2}{*}{Two-stage}&$\toolW_{CC}$&CodeLlama&31.95&33.74&57.18&16.97\\
&&$\toolW_{GC}$&GPT-4o&\textbf{33.81}&\textbf{35.80}&\textbf{62.22}&17.74 \\
\hline
\multirow{5}{*}{GPT-4o} & \multirow{3}{*}{End-to-end}&Instruction&-&62.77&64.11&27.40&34.65\\
&&CoT&-&44.90&45.87&26.95&53.12\\
&&Few-shot Learning&-&48.74&40.36&30.50&49.04\\
\cline{2-8} 
& \multirow{2}{*}{Two-stage}&$\toolW_{GG}$&GPT-4o&63.31&65.47&\textbf{50.08}&14.07 \\

&&$\toolW_{CG}$&CodeLlama&\textbf{67.57}&\textbf{69.81}&48.69&\textbf{12.83} \\
\hline 
\multirow{5}{*}{Claude-3.5} & \multirow{3}{*}{End-to-end}&Instruction&-&53.59&55.47&34.05&24.88\\
&&CoT&-&36.21&37.82&28.45&56.65\\
&&Few-shot Learning&-&53.66&55.40&32.66&35.13\\
\cline{2-8} 
& \multirow{2}{*}{Two-stage}&$\toolW_{ClCl}$&Claude-3.5&51.98&54.26&\textbf{68.94}&27.52 \\

&&$\toolW_{CCl}$&CodeLlama&53.24&\textbf{55.51}&61.23&26.56 \\
\hline 
\end{tabular}

\caption{\label{Reuslt-model}
Evaluation results on the ACPR dataset. All results in the table are reported in percentage (\(\%\)).
}
\vspace{-1em}
\end{table*}

\subsubsection{Baselines.}
To evaluate the effectiveness of our model on the AdaPR task, we build \tool based on popular LLMs, including both closed-source and open-source models. 
For the closed-source baseline, one high-performance model \textbf{GPT-4o}~\cite{openai2024chatgpt4o, achiam2023gpt} and \textbf{Claude-Sonnet-3.5}\footnote{\url{https://www.anthropic.com/news/claude-3-5-sonnet}} are considered. 
For the open-source baseline, we utilize the \textbf{CodeLlama-Instruct-7B}~\cite{roziere2023codellama}, which is a popular foundation model for code-related tasks. 
We use GPT-4o and CodeLlama for stage \Roman{stage} (\textit{i.e.}, bug localization) and stage \Roman{stage2} (\textit{i.e.}, program repair) respectively, denoted as $\boldsymbol{\toolW}_{GC}$, $\boldsymbol{\toolW}_{GG}$ and  $\boldsymbol{\toolW}_{CG}$, $\boldsymbol{\toolW}_{CC}$ respectively. 
All the baselines adopt an end-to-end framework to perform the program repair task. 
Additionally, we incorporate baselines with three widely used LLM-based optimization methods: 
(1) \textbf{Chain-of-Thought (CoT)}: CoT prompting~\cite{kojima2022CoT} elicits complex multi-step reasoning to enhance the model's cognitive capabilities on program repair tasks. 
(2) \textbf{Few-Shot Learning}: Few-shot prompting~\cite{brown2020few-shot} utilizes LLMs' in-context learning abilities to achieve high performance with input-output pairs as extra context. 
(3) \textbf{Fine-Tuning}: LoRA~\cite{hu2021lora} injects trainable rank decomposition matrices into LLMs, updating weights based on supervised labels for the program repair task. 
Further details can be found in the Appendix.

\subsection{Experimental Results} 
\subsubsection{RQ1. Effectiveness Evaluation.}
In this research question, we want to evaluate the effectiveness of our approach on the AdaPR task. 
Table~\ref{Reuslt-model} shows the experimental results of our approach and baselines on our test set. 
It is obvious that: 
(1) Our approach (\textit{e.g.}, $\toolW_{GC}$, $\toolW_{CG}$) outperforms other baselines (\textit{e.g.}, CoT, Few-shot Learning, and Fine-tuning) by a large margin in program repair accuracy, or produces similar results in accuracy while significantly improving consistency. 
The superior performance is due to our Bug Locator's capability to precisely identify the bug locations in the first stage. 
During the first stage, \tool pinpointed the exact buggy line(s) by utilizing the self-debug learning from code-diff samples, enhancing our approach's ability to effectively handle \textit{where to fix} challenge. 
(2) Regarding the program repair consistency, the advantages of our approach over other baselines are obvious. 
For example, the best consistency score achieved by baseline (\textit{i.e.}, CodeLlama Fine-Tuning) is 51.16\%, our $\toolW_{GC}$ achieved a consistency ratio of 62.22\%, significantly outperforming other baseline models. 
We attribute this improved consistency to the effectiveness of the Program Modifier in the second stage. 
During the second stage, we design three key techniques (\textit{i.e.}, Location-Aware Repair learning, Hybrid Training for Selective Reference and Adaptive Preference Learning) to ensure the consistency between the fixed code and buggy code. 
\textbf{Overall, our two-stage framework shows stable and substantial improvements compared to the end-to-end program repair framework, validating the effectiveness of our two-stage approach for the AdaPR task.} 
(3) Additionally, \tool achieves better program repair accuracy when we use GPT-4o in the second stage (\textit{i.e.}, program repair), and achieves better program repair consistency when we use CodeLlama in the same stage. 
This may be because GPT-4o has an advantage over CodeLlama regarding accuracy due to its significantly larger model parameters.
At the same time, CodeLlama has its own strength in identifying bug locations after fine-tuning. 
Therefore, when we choose CodeLlama for the first stage and GPT-4o for the second stage, the optimal performance is obtained.

\begin{table}[t]
\centering
\small
\setlength{\tabcolsep}{0.6mm}
\begin{tabular}{lccc}
\hline 
\multirow{1}{*}{ Model }
& Acc & Improve & consistency \\
\hline
$\boldsymbol{\toolW}_{CC}$& 31.95&33.74&\textbf{57.18} \\
w/o Self-Debug Learning&31.89&33.27&57.06 \\
w/o Location-Aware Repair Learning & 28.05&29.61&54.39\\
w/o Hybrid Training&29.38&31.03&54.35 \\
w/o Adaptive Preference Learning&32.07&33.85&56.78 \\
\hline
$\boldsymbol{\toolW}_{CG}$& \textbf{67.57} &\textbf{69.81}&\textbf{48.69} \\
w/o Self-Debug Learning&65.71&68.11&48.09 \\
\hline
\end{tabular}

\vspace{-0.5em}
\caption{\label{Ablation-study}
Ablation study.
}
\vspace{-1em}
\end{table}

\subsubsection{RQ2. Ablation Study.} 
In this RQ, we conduct an ablation study to assess the contribution of different techniques by systematically removing each component from our approach. 
In particular, for \toolW\textsubscript{CC}, we remove key components (\textit{i.e.}, Self-Debug Learning, Hybrid Training, and Preference Learning) separately. 
For \toolW\textsubscript{CG}, we remove the sole component (\textit{i.e.}, Self-Debug Learning) since GPT-4o is close-sourced. 
The experimental results are illustrated in Table~\ref{Ablation-study}, we can see that: 
(1) Removing Self-Debug Learning, Location-Aware Learning, or Hybrid Training, results in a decline in the performance of accuracy and consistency, which signals the importance and effectiveness of these components.
(2) Although Preference Learning causes a slight decrease in the metric of code accuracy, it further enhances the consistency, indicating this component can effectively reduce modifications during code repair.


\subsubsection{RQ3. Why Our Approach Works/Fails.} 
\begin{figure}[t]
\centering
\includegraphics[width=1.0\linewidth]{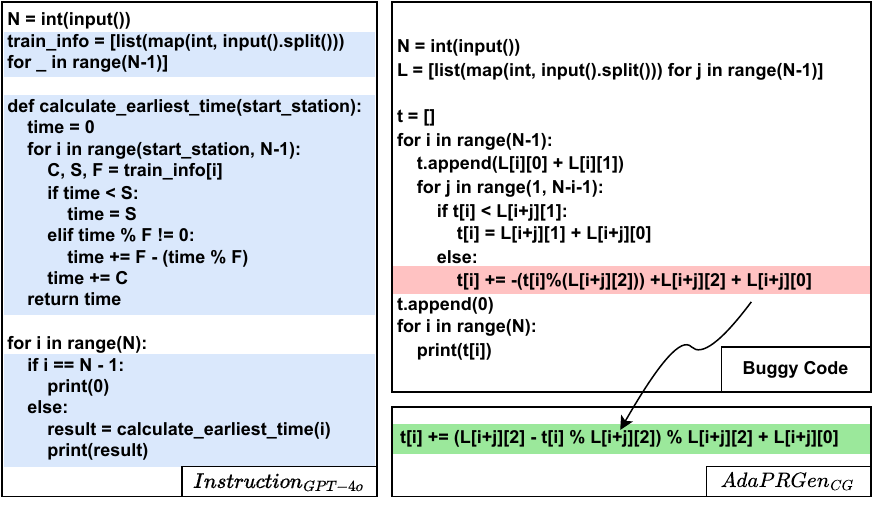}
\vspace{-1.5em}
\caption{The example of adaptive program repair.}
\label{fig:Consistency_Q}
\vspace{-1em}
\end{figure}

We manually inspected test cases where our approach worked and failed. 
Fig.~\ref{fig:Consistency_Q} demonstrates a buggy code fixed by our approach and by GPT-4o. 
Our Bug Locator first precisely identifies the buggy code line and then our program modifier fixes this bug by slightly changing this buggy line. 
While even GPT-4o correctly fixes this bug, GPT-4o modifies the majority of the code lines of the original function, changing the code structure, logic, and semantics of the original code. 
The effectiveness of two-stage framework (Bug Locator + Program Modifier) ensures the correctness and consistency of our generated code patches. 
We also inspected a number of cases where \tool failed to handle. 
We summarize two common failed situations. 
One common failed situation is that the failed test case does not provide sufficient information to precisely identify the bug locations. 
Another bad situation is that the buggy code is too complicated or subtle for \tool to learn. 
For example, complicated bugs may require developers to make code changes across different sub-modules. 
Additional analysis and examples are provided in the Appendix. 

\begin{figure}[t]
\centering
\includegraphics[width=1.0\linewidth]{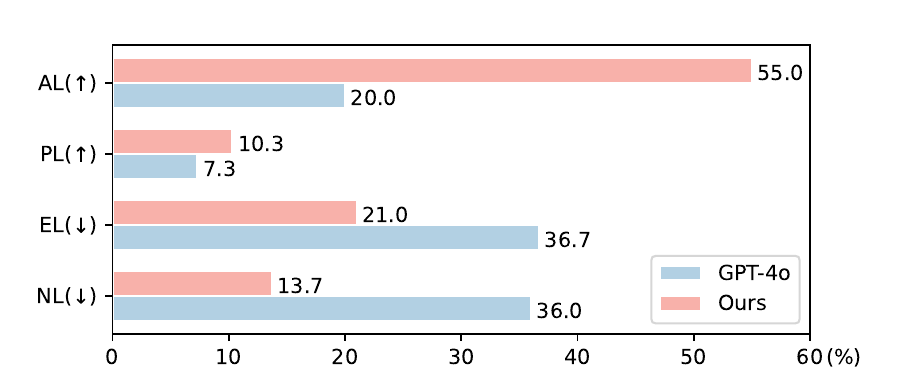}
\vspace{-1.5em}
\caption{The statistical result of the human study.}
\label{fig:Human-study-for-fault-localization}
\vspace{-1em}
\end{figure}
\subsubsection{RQ4. Human Study for Bug Localization.} 
The bug localization in stage one plays an important role for guiding the subsequent program repair process. 
Therefore, in this RQ, we conduct a human study to manually evaluate stage one's performance. 
We compare the bug localization capability of GPT-4o and CodeLlama trained with our framework with human evaluation.  Specifically, 900 samples are provided to 2 experienced evaluators, each evaluator is asked to determine the bug locations independently, the first author is then involved in leading a discussion when they have disagreements. 
Following that, we compare the model-predicted bug locations with human-identified locations in terms of the following aspects:
(1) \textbf{Accurate Localization} (AL) refers to that model-predicted locations match human-identified locations precisely. 
(2) \textbf{Partial Localization} (PL) indicates that only part of model-predicted locations match human-identified locations. 
(3) \textbf{Erroneous Localization} (EL) indicates model-generated locations do not match human-identified locations at all. 
(4) \textbf{No Localization} (NL) denotes that no bugs are identified by models. 
Fig.~\ref{fig:Human-study-for-fault-localization} illustrated the human study results. 
The CodeLlama trained with our framework performs better than GPT-4o in bug localization, with 35.0\% and 3.0\% more examples of AL and PL respectively, while having 15.7\% and 22.3\% fewer EL and NL examples. 
The results show that even GPT-4o fails to identify bug locations of a buggy code effectively, verifying the challenge of this task. 
Our Bug Locator, enables a small-scale LLM (\textit{i.e.}, CodeLlama) to achieve a much superior performance than GPT-4o, validating the effectiveness of our self-debug learning with code diff samples. 

\begin{table}[t]
\centering
\small
\setlength{\tabcolsep}{0.6mm}
\begin{tabular}{lccc}
\hline 
\multirow{1}{*}{ Methods }
& Acc  & Improve &FR\\
\hline
Hybrid Training& \textbf{32.07} &\textbf{33.85}&\textbf{16.73} \\
Supervised Training&27.88&29.76&26.92 \\
Weakly Supervised Training&31.60&33.61&18.94\\
Supervised and Weakly Supervised Training&31.06 &32.23&17.33 \\
\hline
\end{tabular}
\caption{\label{Hybrid-training}
Evaluation results of different training methods.
}
\vspace{-1em}
\end{table}

\subsubsection{RQ5. Hybrid Training Analysis.}
To verify the effectiveness of our hybrid training method for selective reference, in this RQ, we conducted a comparative analysis against other common training methods. 
Particularly, we further designed our experiments to combine both weakly supervised data and supervised data in the training process. As illustrated in Table~\ref{Hybrid-training}, the experimental results show that: (1) Our hybrid training is superior to other training methods in various metrics of the correctness of program repair. 
(2) Compared to supervised training, the correctness of our method has significantly improved, demonstrating the effectiveness of Hybrid Training in avoiding blind modification. 
(3) Compared to supervised and weakly supervised training, the experimental results demonstrate the effectiveness of our hybrid training regardless of the amount of training data.

\section{\textbf{\textsc{Future Work}}}
Several limitations are concerned with our work. 
Firstly, our study is based on Python, which is one of the most popular programming languages used by developers. 
However, our approach is language-independent, we believe our approach can be easily adapted to other programming languages such as C++ or Java.
Secondly, the correctness of the generated code is affected when our model is applied by using adaptive preference learning. 
Exploring effective ways to generate repaired code with reduced modifications while further improving its correctness is an interesting research topic for our future work.

\section{\textbf{\textsc{Conclusion}}}
This research aims to generate fixed code while requiring minimal modifications. 
To perform this novel task, we propose an approach \tool that utilizes self-debug learning to train a Bug Locator to accurately identify bugs and fix code through bug locations. For program repair, we train a Program Modifier through location-aware repair learning. Then we propose hybrid training to effectively avoid blindly modifying incorrect bug locations. Additionally, adaptive preference learning is used to learn fewer modifications. The experimental results
show the effectiveness of our approach for this task. We hope our study lays the foundations for this new
research and provide valuable insights into the potential for bug location and adaptive program repair capabilities of Open-source and closed-source LLMs.
\bibliography{aaai25}

\begin{thebibliography}{38}
\providecommand{\natexlab}[1]{#1}

\bibitem[{Achiam et~al.(2023)Achiam, Adler, Agarwal, Ahmad, Akkaya, Aleman, Almeida, Altenschmidt, Altman, Anadkat et~al.}]{achiam2023gpt}
Achiam, J.; Adler, S.; Agarwal, S.; Ahmad, L.; Akkaya, I.; Aleman, F.~L.; Almeida, D.; Altenschmidt, J.; Altman, S.; Anadkat, S.; et~al. 2023.
\newblock Gpt-4 technical report.
\newblock \emph{arXiv preprint arXiv:2303.08774}.

\bibitem[{Bouzenia et~al.(2023)Bouzenia, Ding, Pei, Ray, and Pradel}]{bouzenia2023tracefixer}
Bouzenia, I.; Ding, Y.; Pei, K.; Ray, B.; and Pradel, M. 2023.
\newblock TraceFixer: Execution trace-driven program repair.
\newblock \emph{arXiv preprint arXiv:2304.12743}.

\bibitem[{Brown et~al.(2020)Brown, Mann, Ryder, Subbiah, Kaplan, Dhariwal, Neelakantan, Shyam, Sastry, Askell et~al.}]{brown2020few-shot}
Brown, T.; Mann, B.; Ryder, N.; Subbiah, M.; Kaplan, J.~D.; Dhariwal, P.; Neelakantan, A.; Shyam, P.; Sastry, G.; Askell, A.; et~al. 2020.
\newblock Language models are few-shot learners.
\newblock \emph{NeurIPS}.

\bibitem[{Chen et~al.(2021)Chen, Tworek, Jun, Yuan, Pinto, Kaplan, Edwards, Burda, Joseph, Brockman et~al.}]{chen2021evaluating}
Chen, M.; Tworek, J.; Jun, H.; Yuan, Q.; Pinto, H. P. D.~O.; Kaplan, J.; Edwards, H.; Burda, Y.; Joseph, N.; Brockman, G.; et~al. 2021.
\newblock Evaluating large language models trained on code.
\newblock \emph{arXiv preprint arXiv:2107.03374}.

\bibitem[{Chen et~al.(2023)Chen, Lin, Sch{\"a}rli, and Zhou}]{chen2023teaching}
Chen, X.; Lin, M.; Sch{\"a}rli, N.; and Zhou, D. 2023.
\newblock Teaching large language models to self-debug.
\newblock \emph{arXiv preprint arXiv:2304.05128}.

\bibitem[{Dissanayake et~al.(2022)Dissanayake, Jayatilaka, Zahedi, and Babar}]{dissanayake2022software}
Dissanayake, N.; Jayatilaka, A.; Zahedi, M.; and Babar, M.~A. 2022.
\newblock Software security patch management-A systematic literature review of challenges, approaches, tools and practices.
\newblock \emph{Information and Software Technology}, 144: 106771.

\bibitem[{Fan et~al.(2023)Fan, Gao, Mirchev, Roychoudhury, and Tan}]{fan2023automated}
Fan, Z.; Gao, X.; Mirchev, M.; Roychoudhury, A.; and Tan, S.~H. 2023.
\newblock Automated repair of programs from large language models.
\newblock In \emph{2023 IEEE/ACM 45th International Conference on Software Engineering (ICSE)}, 1469--1481. IEEE.

\bibitem[{Hu et~al.(2021)Hu, Shen, Wallis, Allen-Zhu, Li, Wang, Wang, and Chen}]{hu2021lora}
Hu, E.~J.; Shen, Y.; Wallis, P.; Allen-Zhu, Z.; Li, Y.; Wang, S.; Wang, L.; and Chen, W. 2021.
\newblock Lora: Low-rank adaptation of large language models.
\newblock \emph{arXiv preprint arXiv:2106.09685}.

\bibitem[{Jiang et~al.(2024)Jiang, Wang, Shen, Kim, and Kim}]{jiang2024survey}
Jiang, J.; Wang, F.; Shen, J.; Kim, S.; and Kim, S. 2024.
\newblock A Survey on Large Language Models for Code Generation.
\newblock \emph{arXiv preprint arXiv:2406.00515}.

\bibitem[{Jiang et~al.(2023)Jiang, Liu, Lutellier, and Tan}]{jiang2023impact}
Jiang, N.; Liu, K.; Lutellier, T.; and Tan, L. 2023.
\newblock Impact of code language models on automated program repair.
\newblock In \emph{2023 IEEE/ACM 45th International Conference on Software Engineering (ICSE)}, 1430--1442. IEEE.

\bibitem[{Jin et~al.(2023)Jin, Shahriar, Tufano, Shi, Lu, Sundaresan, and Svyatkovskiy}]{jin2023inferfix}
Jin, M.; Shahriar, S.; Tufano, M.; Shi, X.; Lu, S.; Sundaresan, N.; and Svyatkovskiy, A. 2023.
\newblock Inferfix: End-to-end program repair with llms.
\newblock In \emph{Proceedings of the 31st ACM Joint European Software Engineering Conference and Symposium on the Foundations of Software Engineering}, 1646--1656.

\bibitem[{Kojima et~al.(2023)Kojima, Shixiang, Reid, Matsuo, and Iwasawa}]{kojima2022CoT}
Kojima, T.; Shixiang, S.; Reid, M.; Matsuo, Y.; and Iwasawa, Y. 2023.
\newblock Large language models are zero-shot reasoners.
\newblock \emph{arXiv preprint arXiv:2205.11916}.

\bibitem[{Krasner(2021)}]{krasner2021cost}
Krasner, H. 2021.
\newblock The cost of poor software quality in the US: A 2020 report.
\newblock \emph{Proc. Consortium Inf. Softw. QualityTM (CISQTM)}, 2.

\bibitem[{Li et~al.(2023)Li, Allal, Zi, Muennighoff, Kocetkov, Mou, Marone, Akiki, Li, Chim et~al.}]{li2023starcoder}
Li, R.; Allal, L.~B.; Zi, Y.; Muennighoff, N.; Kocetkov, D.; Mou, C.; Marone, M.; Akiki, C.; Li, J.; Chim, J.; et~al. 2023.
\newblock Starcoder: may the source be with you!
\newblock \emph{arXiv preprint arXiv:2305.06161}.

\bibitem[{Li et~al.(2022)Li, Choi, Chung, Kushman, Schrittwieser, Leblond, Eccles, Keeling, Gimeno, Dal~Lago et~al.}]{li2022competition}
Li, Y.; Choi, D.; Chung, J.; Kushman, N.; Schrittwieser, J.; Leblond, R.; Eccles, T.; Keeling, J.; Gimeno, F.; Dal~Lago, A.; et~al. 2022.
\newblock Competition-level code generation with alphacode.
\newblock \emph{Science}, 378(6624): 1092--1097.

\bibitem[{Madaan et~al.(2024)Madaan, Tandon, Gupta, Hallinan, Gao, Wiegreffe, Alon, Dziri, Prabhumoye, Yang et~al.}]{madaan2024self}
Madaan, A.; Tandon, N.; Gupta, P.; Hallinan, S.; Gao, L.; Wiegreffe, S.; Alon, U.; Dziri, N.; Prabhumoye, S.; Yang, Y.; et~al. 2024.
\newblock Self-refine: Iterative refinement with self-feedback.
\newblock \emph{Advances in Neural Information Processing Systems}, 36.

\bibitem[{Miceli-Barone et~al.(2023)Miceli-Barone, Barez, Konstas, and Cohen}]{miceli2023larger}
Miceli-Barone, A.~V.; Barez, F.; Konstas, I.; and Cohen, S.~B. 2023.
\newblock The larger they are, the harder they fail: Language models do not recognize identifier swaps in python.
\newblock \emph{arXiv preprint arXiv:2305.15507}.

\bibitem[{Moon et~al.(2023)Moon, Song, Chae, Kang, Kwon, Ong, Hwang, and Yeo}]{moon2023coffee}
Moon, S.; Song, Y.; Chae, H.; Kang, D.; Kwon, T.; Ong, K. T.-i.; Hwang, S.-w.; and Yeo, J. 2023.
\newblock Coffee: Boost your code llms by fixing bugs with feedback.
\newblock \emph{arXiv preprint arXiv:2311.07215}.

\bibitem[{Muennighoff et~al.(2023)Muennighoff, Liu, Zebaze, Zheng, Hui, Zhuo, Singh, Tang, Von~Werra, and Longpre}]{muennighoff2023octopack}
Muennighoff, N.; Liu, Q.; Zebaze, A.; Zheng, Q.; Hui, B.; Zhuo, T.~Y.; Singh, S.; Tang, X.; Von~Werra, L.; and Longpre, S. 2023.
\newblock Octopack: Instruction tuning code large language models.
\newblock \emph{arXiv preprint arXiv:2308.07124}.

\bibitem[{Nam et~al.(2024)Nam, Macvean, Hellendoorn, Vasilescu, and Myers}]{nam2024using}
Nam, D.; Macvean, A.; Hellendoorn, V.; Vasilescu, B.; and Myers, B. 2024.
\newblock Using an llm to help with code understanding.
\newblock In \emph{Proceedings of the IEEE/ACM 46th International Conference on Software Engineering}, 1--13.

\bibitem[{Ni et~al.(2024)Ni, Allamanis, Cohan, Deng, Shi, Sutton, and Yin}]{ni2024next}
Ni, A.; Allamanis, M.; Cohan, A.; Deng, Y.; Shi, K.; Sutton, C.; and Yin, P. 2024.
\newblock NExT: Teaching Large Language Models to Reason about Code Execution.
\newblock \emph{arXiv preprint arXiv:2404.14662}.

\bibitem[{Olausson et~al.(2023)Olausson, Inala, Wang, Gao, and Solar-Lezama}]{olausson2023self}
Olausson, T.~X.; Inala, J.~P.; Wang, C.; Gao, J.; and Solar-Lezama, A. 2023.
\newblock Is Self-Repair a Silver Bullet for Code Generation?
\newblock In \emph{The Twelfth International Conference on Learning Representations}.

\bibitem[{OpenAI(2024)}]{openai2024chatgpt4o}
OpenAI. 2024.
\newblock ChatGPT-4o.
\newblock \url{https://openai.com/index/hello-gpt-4o}.

\bibitem[{Pal et~al.(2024)Pal, Karkhanis, Dooley, Roberts, Naidu, and White}]{pal2024smaug}
Pal, A.; Karkhanis, D.; Dooley, S.; Roberts, M.; Naidu, S.; and White, C. 2024.
\newblock Smaug: Fixing failure modes of preference optimisation with dpo-positive.
\newblock \emph{arXiv preprint arXiv:2402.13228}.

\bibitem[{Papineni et~al.(2002)Papineni, Roukos, Ward, and Zhu}]{papineni2002bleu}
Papineni, K.; Roukos, S.; Ward, T.; and Zhu, W.-J. 2002.
\newblock Bleu: a method for automatic evaluation of machine translation.
\newblock In \emph{Proceedings of the 40th annual meeting of the Association for Computational Linguistics}, 311--318.

\bibitem[{Parreira, Gillet, and Leite(2023)}]{parreira2023robot}
Parreira, M.~T.; Gillet, S.; and Leite, I. 2023.
\newblock Robot Duck Debugging: Can Attentive Listening Improve Problem Solving?
\newblock In \emph{Proceedings of the 25th International Conference on Multimodal Interaction}, 527--536.

\bibitem[{Paul et~al.(2023)Paul, Hossain, Siddiq, Hasan, Iqbal, and Santos}]{paul2023enhancing}
Paul, R.; Hossain, M.~M.; Siddiq, M.~L.; Hasan, M.; Iqbal, A.; and Santos, J. 2023.
\newblock Enhancing automated program repair through fine-tuning and prompt engineering.
\newblock \emph{arXiv preprint arXiv:2304.07840}.

\bibitem[{Puri et~al.(2021)Puri, Kung, Janssen, Zhang, Domeniconi, Zolotov, Dolby, Chen, Choudhury, Decker et~al.}]{puri2021codenet}
Puri, R.; Kung, D.~S.; Janssen, G.; Zhang, W.; Domeniconi, G.; Zolotov, V.; Dolby, J.; Chen, J.; Choudhury, M.; Decker, L.; et~al. 2021.
\newblock Codenet: A large-scale ai for code dataset for learning a diversity of coding tasks.
\newblock \emph{arXiv preprint arXiv:2105.12655}.

\bibitem[{Rafailov et~al.(2024)Rafailov, Sharma, Mitchell, Manning, Ermon, and Finn}]{rafailov2024direct}
Rafailov, R.; Sharma, A.; Mitchell, E.; Manning, C.~D.; Ermon, S.; and Finn, C. 2024.
\newblock Direct preference optimization: Your language model is secretly a reward model.
\newblock \emph{Advances in Neural Information Processing Systems}, 36.

\bibitem[{Roziere et~al.(2023)Roziere, Gehring, Gloeckle, Sootla, Gat, Tan, Adi, Liu, Remez, Rapin et~al.}]{roziere2023codellama}
Roziere, B.; Gehring, J.; Gloeckle, F.; Sootla, S.; Gat, I.; Tan, X.~E.; Adi, Y.; Liu, J.; Remez, T.; Rapin, J.; et~al. 2023.
\newblock Code llama: Open foundation models for code.
\newblock \emph{arXiv preprint arXiv:2308.12950}.

\bibitem[{Shahriar and Zulkernine(2012)}]{shahriar2012mitigating}
Shahriar, H.; and Zulkernine, M. 2012.
\newblock Mitigating program security vulnerabilities: Approaches and challenges.
\newblock \emph{ACM Computing Surveys (CSUR)}, 44(3): 1--46.

\bibitem[{Sobania et~al.(2023)Sobania, Briesch, Hanna, and Petke}]{sobania2023analysis}
Sobania, D.; Briesch, M.; Hanna, C.; and Petke, J. 2023.
\newblock An analysis of the automatic bug fixing performance of chatgpt.
\newblock In \emph{2023 IEEE/ACM International Workshop on Automated Program Repair (APR)}, 23--30. IEEE.

\bibitem[{Xia, Wei, and Zhang(2023)}]{xia2023automated}
Xia, C.~S.; Wei, Y.; and Zhang, L. 2023.
\newblock Automated program repair in the era of large pre-trained language models.
\newblock In \emph{2023 IEEE/ACM 45th International Conference on Software Engineering (ICSE)}, 1482--1494. IEEE.

\bibitem[{Xia and Zhang(2022)}]{xia2022less}
Xia, C.~S.; and Zhang, L. 2022.
\newblock Less training, more repairing please: revisiting automated program repair via zero-shot learning.
\newblock In \emph{Proceedings of the 30th ACM Joint European Software Engineering Conference and Symposium on the Foundations of Software Engineering}, 959--971.

\bibitem[{Yao et~al.(2022)Yao, Zhao, Yu, Du, Shafran, Narasimhan, and Cao}]{yao2022react}
Yao, S.; Zhao, J.; Yu, D.; Du, N.; Shafran, I.; Narasimhan, K.; and Cao, Y. 2022.
\newblock React: Synergizing reasoning and acting in language models.
\newblock \emph{arXiv preprint arXiv:2210.03629}.

\bibitem[{Ye et~al.(2022)Ye, Martinez, Luo, Zhang, and Monperrus}]{ye2022selfapr}
Ye, H.; Martinez, M.; Luo, X.; Zhang, T.; and Monperrus, M. 2022.
\newblock Selfapr: Self-supervised program repair with test execution diagnostics.
\newblock In \emph{Proceedings of the 37th IEEE/ACM International Conference on Automated Software Engineering}, 1--13.

\bibitem[{Zhang et~al.(2022)Zhang, Panthaplackel, Nie, Li, and Gligoric}]{zhang2022coditt5}
Zhang, J.; Panthaplackel, S.; Nie, P.; Li, J.~J.; and Gligoric, M. 2022.
\newblock Coditt5: Pretraining for source code and natural language editing.
\newblock In \emph{Proceedings of the 37th IEEE/ACM International Conference on Automated Software Engineering}, 1--12.

\bibitem[{Zhang et~al.(2023)Zhang, Li, Li, Li, and Jin}]{zhang2023self}
Zhang, K.; Li, Z.; Li, J.; Li, G.; and Jin, Z. 2023.
\newblock Self-edit: Fault-aware code editor for code generation.
\newblock \emph{arXiv preprint arXiv:2305.04087}.

\end{thebibliography}
\appendix
\section{Experimental Setups}
\subsubsection{Dataset}
\paragraph{Dataset Details of ACPR}
We construct the first dataset, named ACPR (Accuracy-Consistency Program Repair) for AdaPR (Adaptive Program Repair) task. we evaluate correctness through unit tests.
CodeNet~\cite{puri2021codenet} includes an average of 4 test cases per problem.
To improve coverage, we include additional test cases from AlphaCode~\cite{li2022competition}. To ensure the correctness and effectiveness of judging, we exclude the submission if their code presents inconsistency in judging results based on existing test cases with the result in competitions.
Since validation of some error types is extremely time-consuming (\textit{e.g.}, Time limit Exceeded), the buggy code focuses on logical errors in our research.
The correctness of a program is checked by executing it on the test cases and comparing the program
output with the expected correct output.

Our approach incorporates a set of strategies to ensure ``the correct submission is based on the failed submission''. Specifically, for submissions of the same programmer on the same problem, the following steps are performed: (1) \textbf{Submission Order}: In our data processing, we strictly require that the correct version be submitted after the incorrect one.
(2) \textbf{Code Cleaning}: To concentrate more precisely on logical changes in the code, we removed comments before calculating similarity. This step ensures that similarity between code pairs reflects functional and logical changes.
(3) \textbf{Similarity-Based Code Filtering}: We used BLEU~\cite{papineni2002bleu} scores to compute the textual similarity between the correct and incorrect submissions. Based on observations from a subset of data, we established a similarity threshold of 0.6. Only code pairs that exceed this threshold are considered as valid and retained as suitable training data.
(4) \textbf{Maximum Similarity Selection}: Among all candidate pairs that meet the aforementioned similarity threshold, we only selected the pair with the highest BLEU score as the final pair for training.

These steps are taken to ensure the training data aligns with our assumption (\textit{i.e.}, the correct submission fixes the bug based on the incorrect submission).

\paragraph{Data construction of Bug Locations}
We construct data of bug locations in buggy code using in form of \texttt{Code Diff}. Specifically, we use the ``git diff'' command to compare pairs of buggy programs and corresponding passed programs for the same problem in the dataset. Each pair of the buggy program and the passed program is written into two separate files, and the ``git diff'' command is used to identify the differences between the two programs to achieve bug locations. Additionally, we process the diff-file as follows: (1) We delete the header information from the diff content, which refers to the initial lines of the git diff output that include the file names, indexes, and line number ranges. (2) For the differing content, we retain the lines of code starting with `-' and remove the lines starting with `+'.
\paragraph{Data construction of Program Trace Information}
For each buggy program, we randomly select a failed test case to serve as the input. Using Python's traceback module\footnote{\url{https://docs.python.org/3/library/traceback.html}} as a debug tool, we obtain the trace information for each line of code: (1) The order of execution for the line. (2) The variables changed after execution. Each line's trace information is formatted as an inline comment and embedded within the original code to maintain the code structure. 

Considering the token limit for LLMs' training, we apply several special treatments to the program trace information: (1) For loop execution information, when the number of iterations exceeds three, we replace the intermediate iteration variables with ellipses. (2) We remove the values of function objects (\textit{e.g.}, excluding information in the form of ``f\_name=\textless function f\_name at 0x...\textgreater''). (3) For one-dimensional array objects with more than 20 elements, we retain only the first and last two values in the trace information, with the omitted parts replaced by ellipses. (4) For multi-dimensional array objects, only the array name is preserved in the trace information.

\subsection{Implementation Details.}
We choose CodeLlama-instruction-7B as the base LLM. In the first training stage, we employed the AdamW optimizer with a learning rate set to 5e-5.
The learning rate schedule was managed using the WarmupDecayLR scheduler, where the total number of steps was 100, the initial learning rate at the start of the warm-up phase was 0.0, the peak learning rate reached at the end of the warm-up phase was 5e-5.
The batch size is 16. 
In the second training stage, we employed the AdamW optimizer with a learning rate set to 5e-6, using the cosine scheduler. The warm-up process included 100 steps. The batch size is 16. we set $\beta$  as 0.1 and $\lambda$ as 5.
During decoding, the diff-based file and fixed code are generated using greedy decoding. 
\subsubsection{Metrics of Code Consistency.}  
To evaluate the consistency between the post-modified fixed code and the premodified buggy code, we propose metrics \textbf{Code Consistency Rate (Consistency)}: It calculates the proportion of lines of code that are preserved after modification. 
It is defined as follows:
\begin{equation}
    Consistency=\frac{r}{k},
\end{equation}
where $k$ indicates the total number of code lines in the fixed code, and $r$ indicates the number of code lines preserved in the after-modification code. We use  Git\footnote{\url{https://git-scm.com/}} to calculate the consistency.
Specifically, by using the ``git diff'' command to compare the fixed code and the buggy code, we obtain the number of lines of code that were deleted or changed, denoted as $a$, and the number of lines of code that were added or changed, denoted as $b$.
The consistency index can be further expressed as the following formula:
\begin{equation}
    Consistency=\frac{k-a}{k+(b-a)},
\end{equation}
where $Consistency \in [0,1]$, reflects the evaluation of consistency from the perspective of retention and addition.

For the ACC and Improve metrics, the code generated by the closed-source LLMs often does not follow the requirements of the AdaPR task and directly produces correct code. In such cases, even if the generated code is correct, we do not consider it as a contribution. The final results are determined through evaluations by three experienced evaluators.

As for the Consistency metric, open-source LLMs often do not modify buggy code, resulting in a consistency score of 1.0. For such cases, we set the consistency score to 0. We use Git tools to check whether any lines have been deleted or modified to determine if a change has been made.

\subsubsection{Baselines.} 
As shown in Fig.~\ref{fig:instruction-prompt}, We utilize a specific instruction prompt to measure the capabilities of GPT-4o~\cite{openai2024chatgpt4o, achiam2023gpt} and CodeLlama-Instruct-7B~\cite{roziere2023codellama} on adaptive program repair task.  
For the prompt template, $\text{\textless language\textgreater}$ is filled with the type of language.  
We further deploy several widely used LLM-based optimization methods on GPT-4o and CodeLlama-Instruct-7B to evaluate their performance: (1) \textbf{Chain-of-Thought (CoT)}~\cite{kojima2022CoT}: we measure the adaptive program repair skills of GPT-4o and CodeLlama-Instruct-7B by utilizing CoT prompting method and asking LLMs to think step by step. The prompt template we used is shown in Fig.~\ref{fig:cot-prompt}. (2) \textbf{Few-Shot Learning}~\cite{brown2020few-shot}: we add two examples to the prompt as extra conditions to enhance model performance. Each example contains three parts, including the description of the programming task, the buggy code, and the corresponding fixed code. The few-shot prompting is shown in Fig.~\ref{fig:few-shot-prompt}. (3) \textbf{Fine-Tuning}: LoRA~\cite{hu2021lora} is an effective way to increase the model's performance in specific fields. Since GPT-4o is closed-source, we only fine-tuned CodeLlama-Instruct-7B by LoRA. During training, the prompt template is the same as in Fig.~\ref{fig:instruction-prompt}.

Additionally, for the generation of GPT-4o, we apply top-$p$ sampling and temperature, where p = 0.7 and T = 1.0, and the number of generation tokens is limited to 2048. As for the CodeLlama, the decoding strategy is set to a greedy strategy. 

\begin{figure}[!thb]
    \centering
        \fcolorbox{black}{gray!6}{%
            \parbox{0.98\linewidth}{%
                \small
                \noindent$\bullet$ \textbf{Instruction}: Given a programming question and a corresponding piece of buggy code written in \textless language\textgreater, please correct the code by modifying the provided buggy code.\\
                \rule{\linewidth}{0.48pt}
                $\bullet$ \textbf{Programming Task}: $q$\\
                $\bullet$ \textbf{Buggy Code}: $c$
            }
        }
    \caption{Instruction prompting for LLMs.}
    \label{fig:instruction-prompt}
\end{figure}

\begin{figure}[!thb]
    \centering
        \fcolorbox{black}{gray!6}{%
            \parbox{0.98\linewidth}{%
                \small
                \noindent$\bullet$ \textbf{Instruction}: Given a programming question and a corresponding piece of buggy code written in \textless language\textgreater, please correct the code by modifying the provided buggy code. Let's think step by step.\\
                \rule{\linewidth}{0.48pt}
                $\bullet$ \textbf{Programming Task}: $q$\\
                $\bullet$ \textbf{Buggy Code}: $c$
            }
        }
    \caption{Chain-of-Thought prompting for LLMs.}
    \label{fig:cot-prompt}
\end{figure}

\begin{figure}[!thb]
    \centering
        \fcolorbox{black}{gray!6}{%
            \parbox{0.98\linewidth}{%
                \small
                \noindent$\bullet$ \textbf{Instruction}: Given a programming question and a corresponding piece of buggy code written in \textless language\textgreater, please correct the code by modifying the provided buggy code. Here are examples of program repair:\\
                \vspace{0.2em}
                \rule{\linewidth}{0.48pt}
                $\bullet$ \textbf{Example 1}:\\
                \vspace{-0.1em}
                \hspace*{1em}\textbf{Programming Task}: $q_1$\\
                \vspace{-0.1em}
                \hspace*{1em}\textbf{Buggy Code}: $c_1$\\
                \vspace{-0.1em}
                \hspace*{1em}\textbf{Corrected Code}: $c'_1$\\
                $\bullet$ \textbf{Example 2}:\\
                \vspace{-0.1em}
                \hspace*{1em}\textbf{Programming Task}: $q_2$\\
                \vspace{-0.1em}
                \hspace*{1em}\textbf{Buggy Code}: $c_2$\\
                \vspace{-0.1em}
                \hspace*{1em}\textbf{Corrected Code}: $c'_2$\\
                \rule{\linewidth}{0.48pt}
                $\bullet$ \textbf{Programming Task}: $q$\\
                $\bullet$ \textbf{Buggy Code}: $c$
            }%
        }
    \caption{Few-shot prompting for LLMs.}
    \label{fig:few-shot-prompt}
\end{figure}

\begin{figure*}[t]
    \centering
    \includegraphics[width=\textwidth]{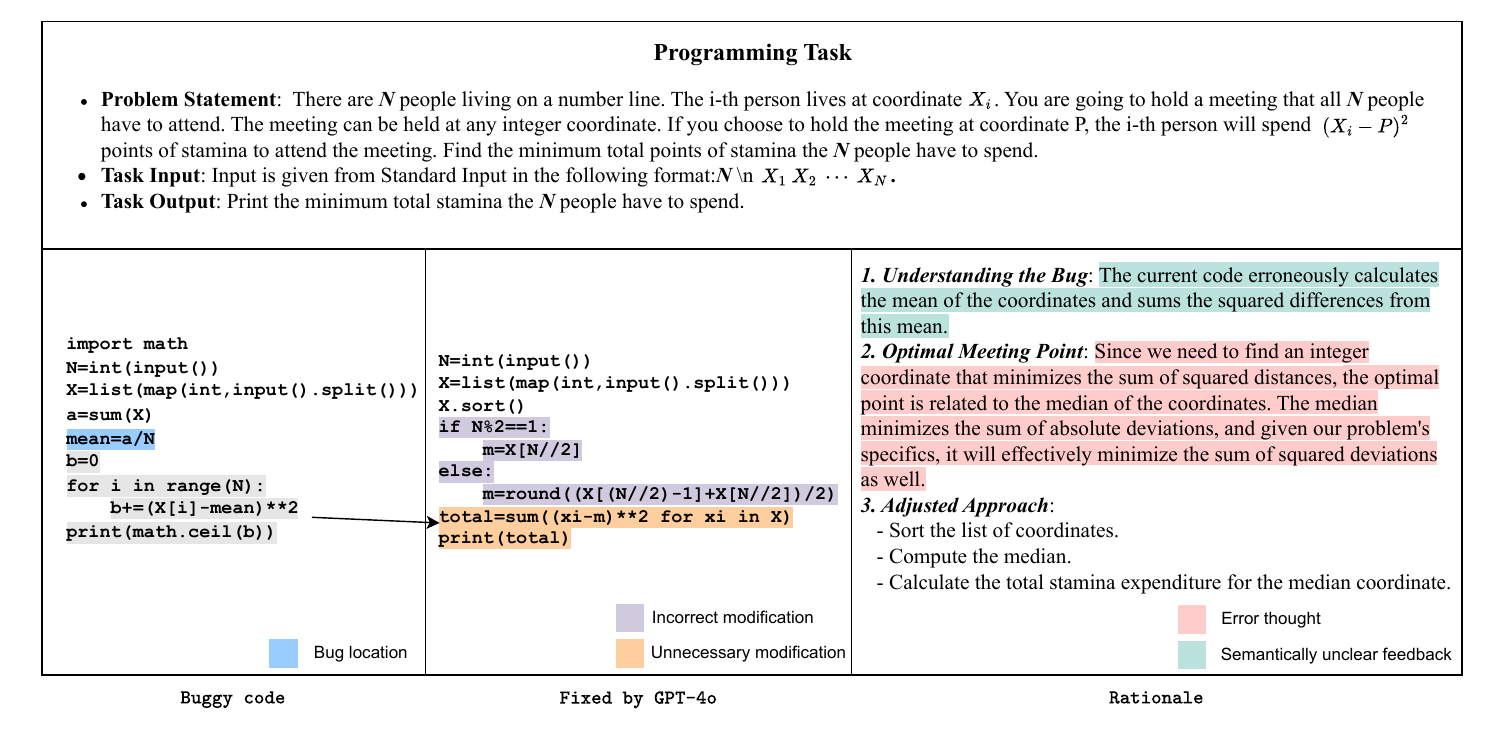}
    \caption{An example of misleading repair caused by CoT.}
    \label{fig:cot}
\end{figure*}

\begin{figure*}[t]
    \centering
    \includegraphics[width=\textwidth]{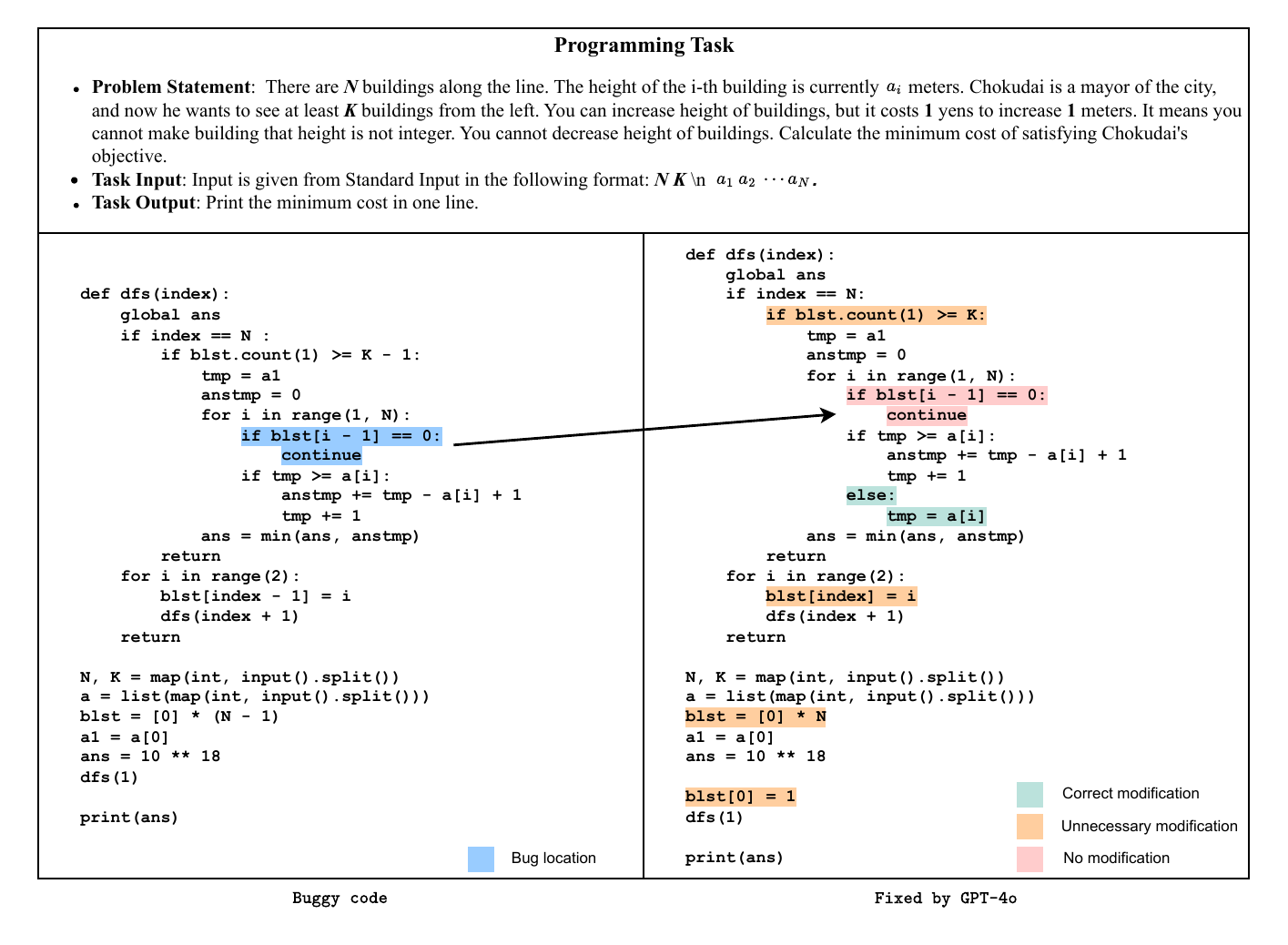}
    \caption{An example of difficulties in locating and fixing bugs by Few-Shot Learning.}
    \label{fig:few-shot}
\end{figure*}

\section{Experimental Results}
\subsection{RQ1. Effectiveness Evaluation.} 
\subsubsection{Qualitative Analysis of Baseline Results.} 
In this section, we analyze the generated results after applying several baseline methods. This study explores drawbacks and challenges in applying these methods for adaptive program repair, including semantically unclear natural language feedback, the introduction of misleading information, and difficulties in avoiding unnecessary modifications.

(1) \textbf{Chain-of-Thought}: When utilizing the CoT method for adaptive program repair, we discover that relying solely on CoT may mislead the entire program repair process and result in unnecessary modifications and poor outcomes due to semantically unclear feedback and incorrect reasoning during the thought process. 
A representative example is shown in Fig.~\ref{fig:cot}, with GPT's thought process detailed in the rationale section. 
This example highlights the issues of semantic ambiguity and the introduction of misleading information as follows: 
(i) \textbf{Semantic ambiguity leads to unnecessary modifications.} As illustrated in Fig.~\ref{fig:cot} GPT-4o's understanding of the bugs is semantically ambiguous, which prevents accurately pinpointing bugs. Firstly, GPT-4o fails to explicitly point out that ``mean=a/N'' is incorrect. Secondly, in its natural language description, it states that there are two bugs in calculating the mean of the coordinates and summing the squared differences. Actually, the summation operation considered to be erroneous by GPT-4o is correct. Consequently, the semantically unclear analysis leads to incorrect bug localization, resulting in unnecessary modifications (\textit{i.e.}, colored in yellow). 
(ii) \textbf{Misleading reasoning results in a poor repair outcome.} In the optimal meeting point section of Fig.~\ref{fig:cot}, GPT-4o incorrectly attributes the bug to choosing the mean as the ``meeting point'' instead of the median. Actually, the bug is in calculating the mean without ensuring the ``meeting point'' is an integer. The incorrect reasoning and misleading information lead GPT-4o to replace the mean calculation with the median calculation, resulting in an error.

(2) \textbf{Few-Shot Learning}: A few adaptive program repair examples are provided as context through few-shot prompting to help GPT-4o understand how to perform repair tasks. These examples in natural language form are not semantically clear enough for GPT-4o to identify and repair bugs effectively as follows:
(i) \textbf{Examples cannot directly help to locate bugs accurately and avoid unnecessary modifications.} As shown in Fig \ref{fig:few-shot}, GPT-4o repairs one bug in the buggy code (\textit{i.e.}, colored in green): it adds the missing branch condition for ``tmp\textless a[i]''. However, it leaves another unresolved (\textit{i.e.}, colored in red): it doesn't update the variable ``tmp'' of the height, in the case where ``blst[i-1]==0''. Additionally, it treats several lines of code that are correct as bugs and makes some unnecessary modifications (\textit{i.e.}, colored in yellow). This situation indicates that GPT-4o has made a confused fault localization which also brings difficulties to the task of reducing modifications while repairing. One reason for the confused fault localization is that the in-context examples of repair cannot directly point out the bugs in the current buggy code. (ii) \textbf{Examples don't enable LLMs to repair accurately.} GPT-4o doesn't repair the bugs correctly, as the modified lines (\textit{i.e.}, colored in yellow) introduce a new bug (\textit{i.e.}, colored in red) to the code: the ``blst[i-1]'' no longer indicates whether the $i$-th building is visible. The introduction of new bugs is mainly because the examples only show how to repair their own bugs, offering little useful information for repairing the current buggy code.

\subsubsection{Qualitative Analysis of AdaPatcher.}
\begin{figure*}[t]
    \centering
    \includegraphics[width=\textwidth]{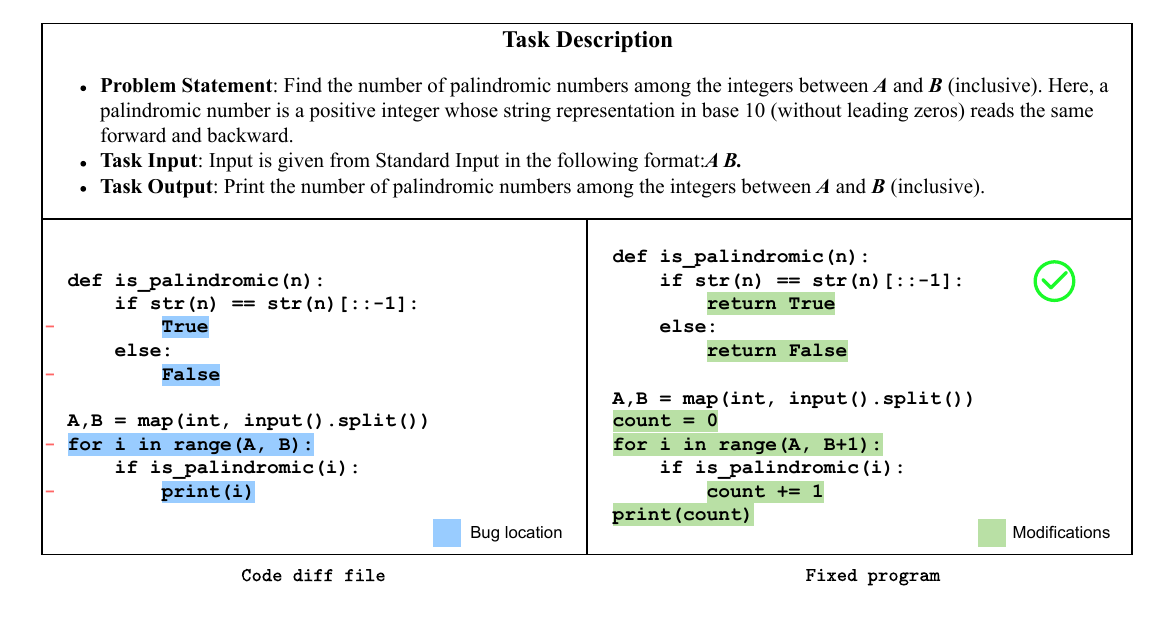}
    \caption{An example to illustrate how \toolW\xspace works.}
    \label{fig:qa_two_stage}
\end{figure*}

To further illustrate our two-stage approach \toolW, as shown in Fig.~\ref{fig:qa_two_stage}, we utilize an example of program repaired by \toolW\textsubscript{CC} to explain the workflow of \toolW\xspace during inference: (1) In the first stage, the Bug Locator accurately identifies the bugs in the buggy code and marks the buggy lines with ``-''. (2) In the second stage, the Program Modifier performs targeted program fixes based on the bug locations, effectively reducing modifications during the repair process.

\begin{figure*}[t]
    \centering
    \includegraphics[width=\linewidth]{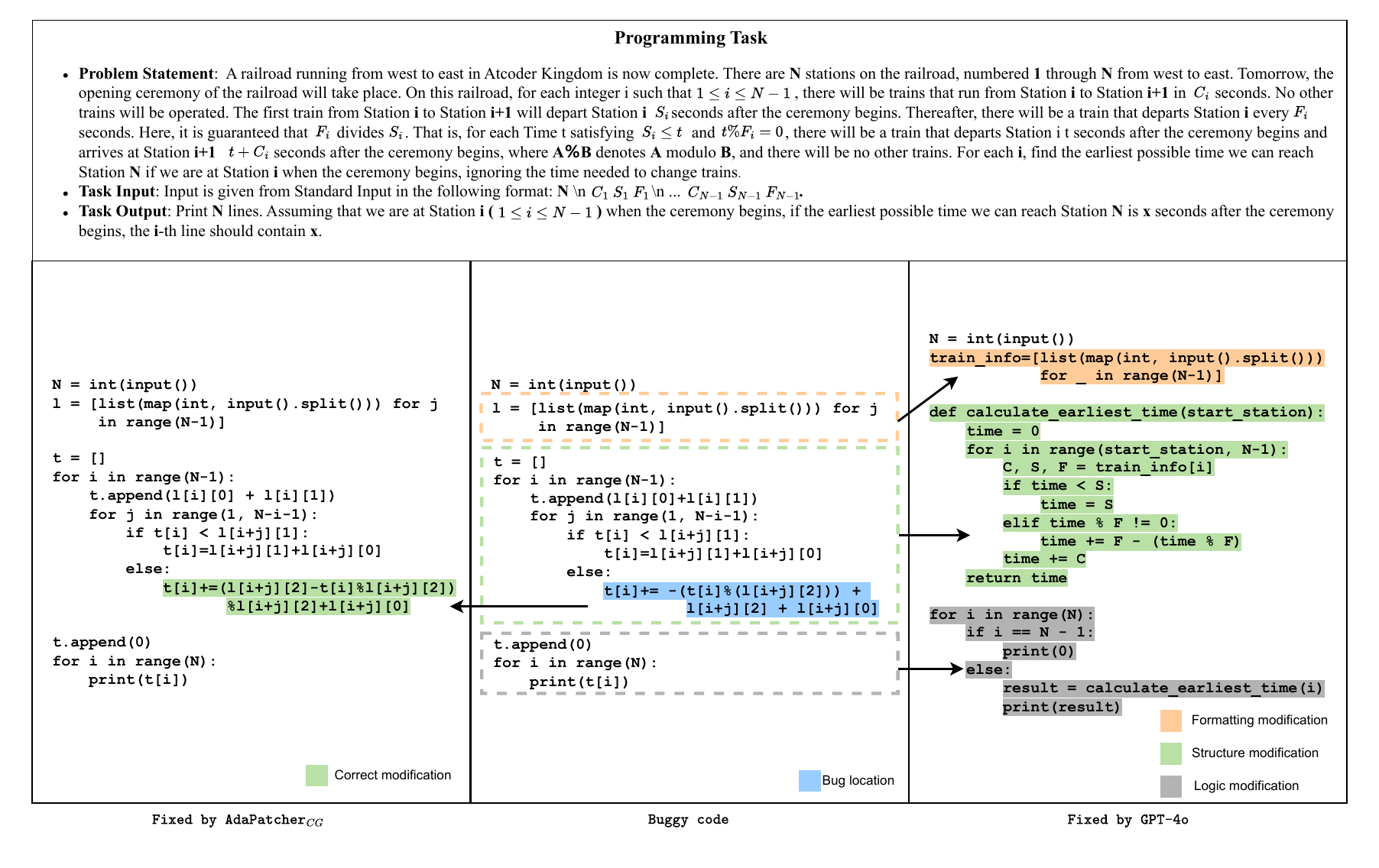}
    \caption{The fixed code by Instruction\textsubscript{GPT-4o} and {\toolW\textsubscript{CG}}.}
    \label{fig:GPT-GC}
\end{figure*}

\begin{figure*}[t]
    \centering
    \includegraphics[width=\linewidth]{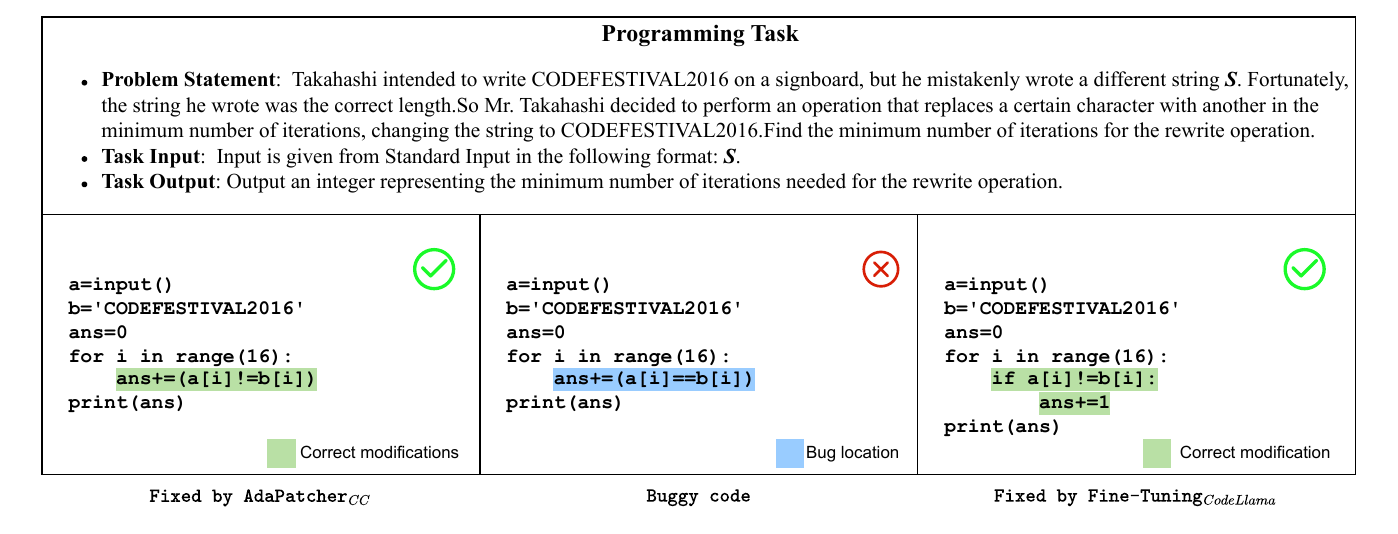}
    \caption{The repaired code by Fine-Tuning\textsubscript{CodeLlama} and {\toolW\textsubscript{CC}}}
    \label{fig:FT-CC}
\end{figure*}

\begin{figure*}[t]
    \centering
    \includegraphics[width=0.98\textwidth]{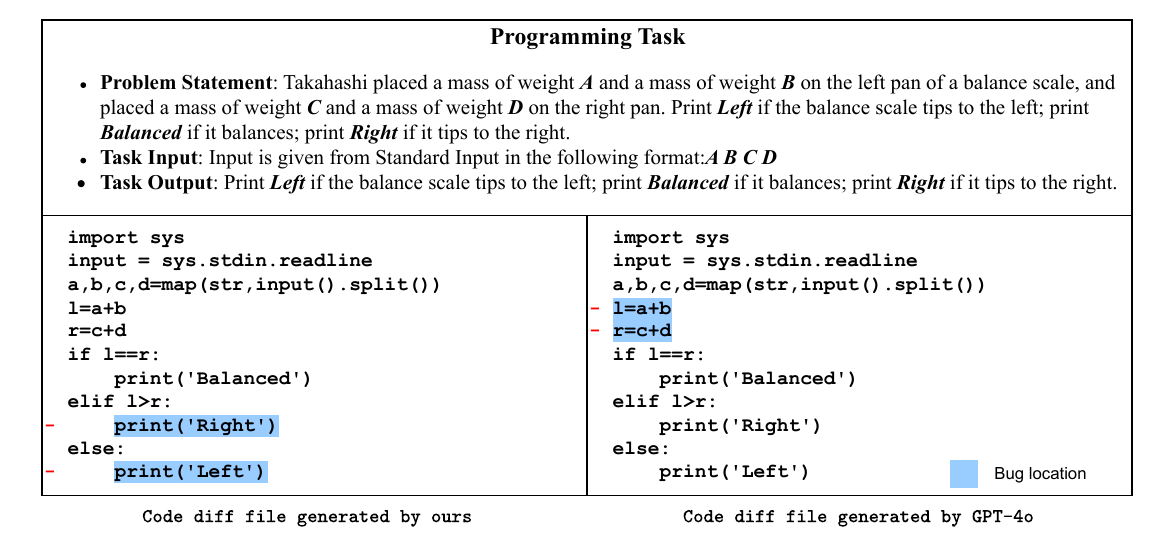}
    \caption{The comparison of accurate semantic understanding between GPT-4o and Ours.}
    \label{fig:qa-of-fl-2}
\end{figure*}
\subsection{RQ3. Why Our Approach Works/Fails.}
\subsubsection{Qualitative Analysis of the Effectiveness.}
We evaluate Instruction\textsubscript{GPT-4o} vs. \toolW \textsubscript{CG} and Fine-Tuning\textsubscript{CodeLlama} vs. \toolW\textsubscript{CC} to demonstrate that our two-stage approach \tool has a significant effect in repair accuracy while ensuring reducing modifications. 

(1) \textbf{Instruction\textsubscript{GPT-4o} vs. AdaPatcher\textsubscript{CG}}: 
We compare the repair results of Instruction\textsubscript{GPT-4o} and \toolW\textsubscript{CG}, showing that our approach can effectively reduce modifications and maintain consistency in terms of code formatting, logic, and structure, as shown in Fig.~\ref{fig:GPT-GC}. 
We analyze this example in further detail:
(i) \textbf{The code fixed by \toolW\textsubscript{CG} is more consistent in terms of code formatting}. The consistency in code formatting is mainly reflected in the naming of variables. In the second line of the original buggy code, the variable ``l'' is used to receive the input, and the intermediate variable ``j'' is used for an iteration. In the fixed code of Instruction\textsubscript{GPT-4o}, even though the same logic is adapted to receive input, it still rewrites the variable name ``l'' to ``train\_info'' and ``j''  to ``\_'' (\textit{i.e.}, colored in yellow). In contrast, \toolW\textsubscript{CG} perfectly inherits and retains these variable names.
Similarly, Instruction\textsubscript{GPT-4o} changes the variable ``t'' to ``time'', whereas \toolW\textsubscript{CG} retains the original naming. 
Instruction\textsubscript{GPT-4o}'s behavior of renaming variables disrupts the consistency between the buggy code and the fixed code in terms of code formatting. In contrast, \toolW\textsubscript{CG} under our method better maintains consistency of the code formatting and avoids unnecessary modifications.
(ii) \textbf{The code fixed by \toolW\textsubscript{CG} is more consistent in terms of code logic}: The consistency in code logic is mainly reflected in the program's execution logic and control flow. In the buggy code, all results are computed before being output sequentially (indicated by the gray box). 
As shown in the Instruction\textsubscript{GPT-4o}'s code, significant modifications are made to the program's execution logic and control flow, specifically by incorporating ``if-else'' statements to compute and output the results one by one (\textit{i.e.}, colored in gray). In contrast, \toolW\textsubscript{CG} maintains consistency with the buggy code in terms of execution logic and control flow.
(iii) \textbf{The code fixed by \toolW\textsubscript{CG} is more consistent in terms of code structure}: The consistency in code structure is mainly reflected in the encapsulation of functionalities. In the buggy code, the main process for calculating results (indicated by the green box) is not encapsulated. Instruction\textsubscript{GPT-4o}, however, encapsulates the functionality (\textit{i.e.}, colored in green), resulting in significant structural changes. In contrast, \toolW\textsubscript{CG} does not alter the code structure of the functionality and corrects the erroneous line in the code.

(2) \textbf{Fine-Tuning\textsubscript{CodeLlama} vs. {AdaPatcher\textsubscript{CC}}}: 
We compare the generated results generated by Fine-Tuning\textsubscript{CodeLlama} and {\toolW\textsubscript{CC}}, which further demonstrates that our approach can reduce modifications and ensure correct fixes. 
As shown in Fig.~\ref{fig:FT-CC}, we choose a representative pair of fixed codes to illustrate this conclusion. 
From the fixed code, it can be observed that both Fine-Tuning\textsubscript{CodeLlama} and {\toolW\textsubscript{CC}} successfully repair the buggy code using their respective methods. Although the code fixed by Fine-Tuning\textsubscript{CodeLlama} is overall similar to the buggy program, there is still a certain degree of inconsistency in terms of code structure: the buggy program implicitly computes the value of ``ans'' using a ternary expression, whereas the Fine-Tuning\textsubscript{CodeLlama} explicitly exposes this computation in its repaired program. In this regard, our model's repaired program is more structurally similar to the buggy program, which reflects that {\toolW\textsubscript{CC}} can effectively reduce modifications while ensuring correct fixes.

\begin{figure*}[ht]
    \centering
    \includegraphics[width=\linewidth]{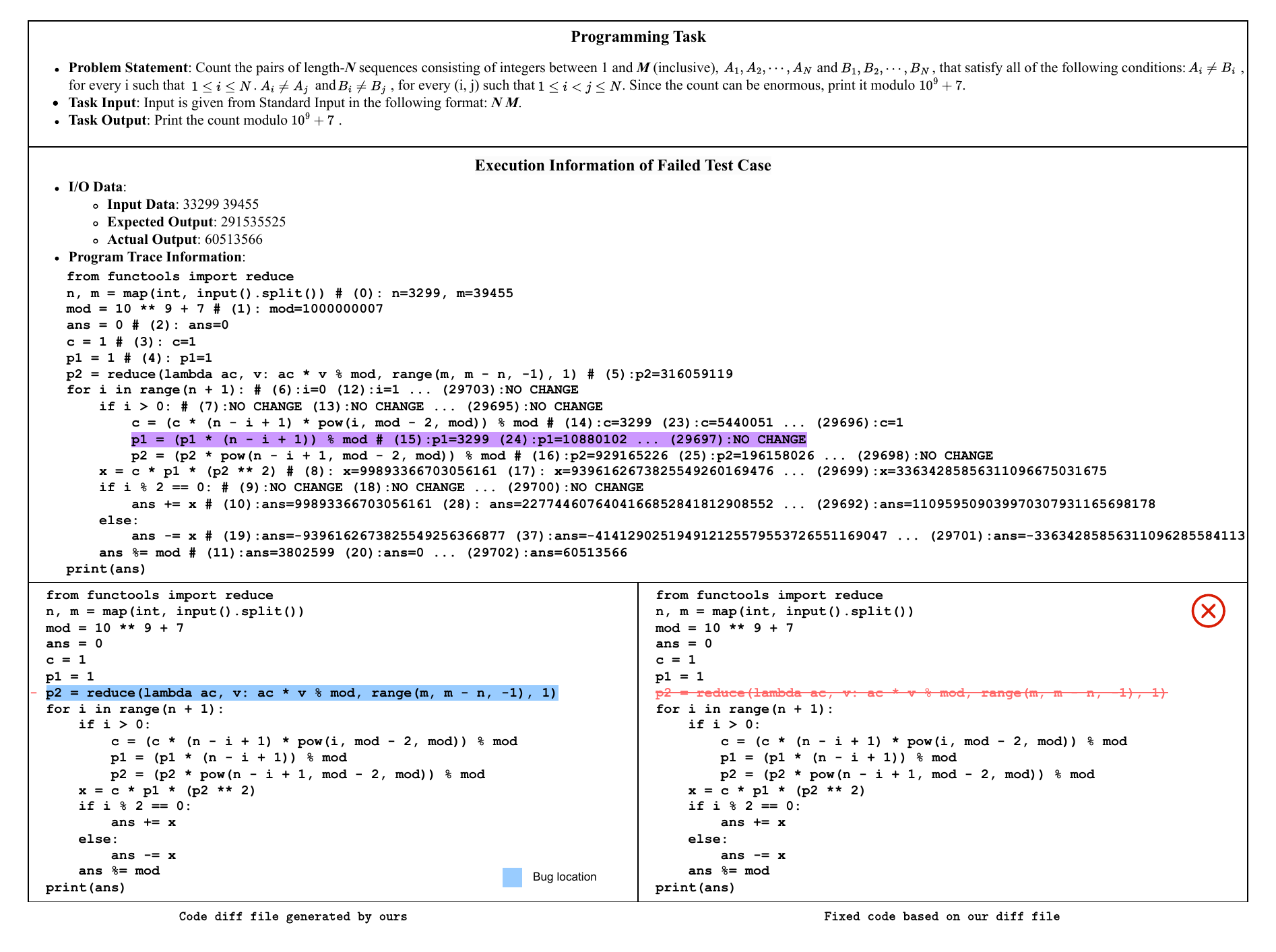}
    \caption{An example of insufficient information causing the failure in fixing.}
    \label{fig:insufficient_information}
\end{figure*}

\subsubsection{Qualitative Analysis of Failure Cases.}
\begin{figure*}[t]
    \centering
    \includegraphics[width=\linewidth]{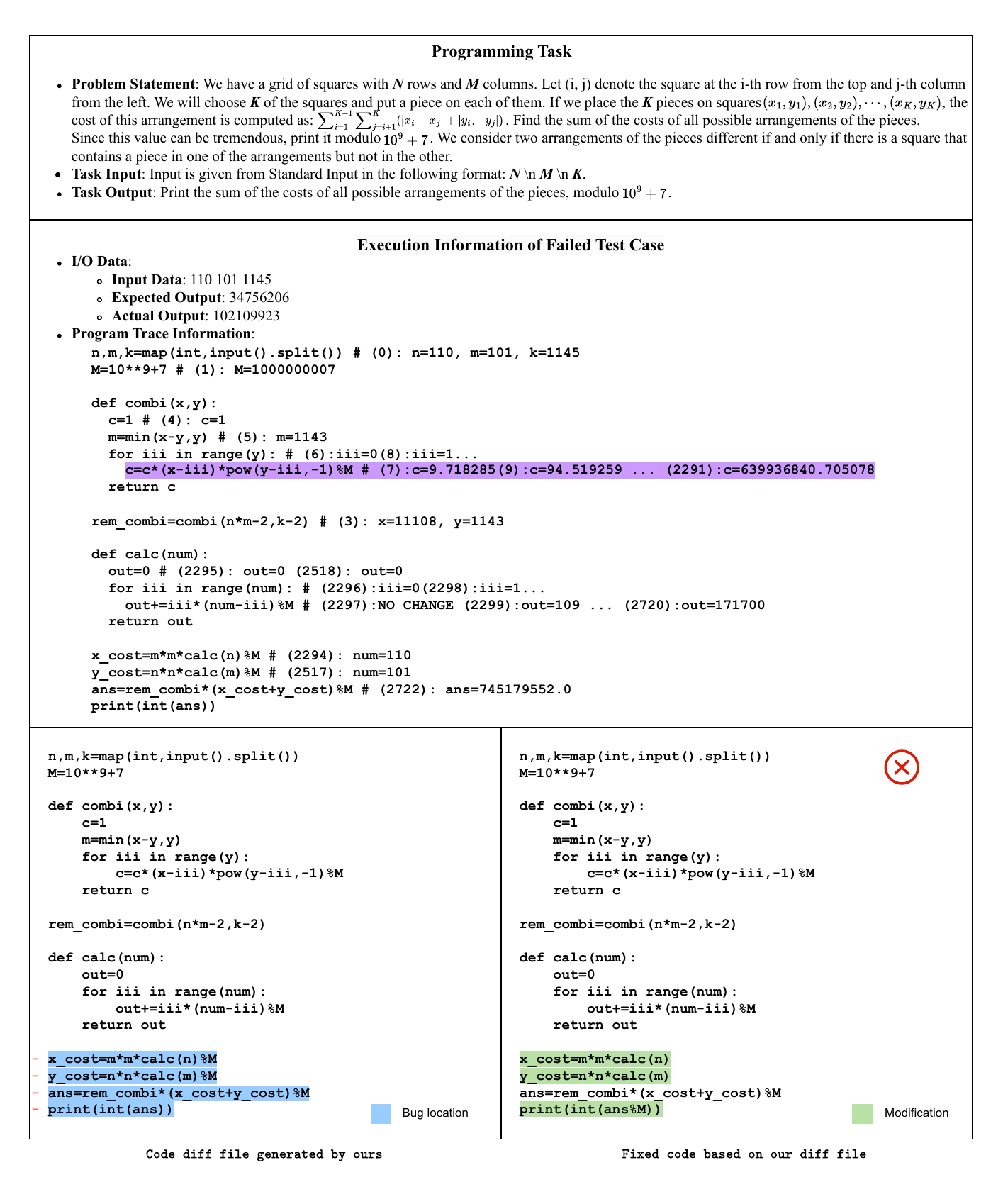}
    \caption{An example of a complex code structure causing the failure in fixing.}
    \label{fig:complicated_code}
\end{figure*}

To further illustrate why our two-stage approach \toolW\xspace fails to fix some buggy programs, we summarize two common failed situations. For each situation, we select an example fixed by \toolW\textsubscript{CC} for the detailed explanation.
(1) \textbf{The failed case does not provide sufficient information to precisely identify the bug locations.} Though we incorporate a hybrid training strategy in Program Modifier to reduce the negative impact brought by the incorrect bug locations, the correctness of the bug locations remains closely related to the repair results. The reason for generating incorrect bug locations is that the failed case does not provide sufficient information. 
As shown in Fig.~\ref{fig:insufficient_information}, intuitively, it is hard for the Bug Locator to pinpoint bugs based on the failed case.
Even if we highlight the buggy line with a purple background in the execution information of the failed test case, it is still challenging for humans to determine from the provided information that this particular line of code is an error, let alone for the Bug Locator.
The underlying reason is that the information provided by the failed case is limited, making it difficult for Bug Locator to accurately identify the error line of code. Consequently, it incorrectly marks other correct statements as a bug, leading to an ineffective repair process.
(2) \textbf{The buggy code is too complicated or subtle for \toolW\xspace to learn.} We find that some buggy codes are too difficult for Bug Locator to locate bugs, especially those programs with complex calling structures. When the bug occurs within a custom function in the program, it often implies that the returned value of this function is also erroneous. Our Bug Locator, based on the erroneous return value of the calling function statement, sometimes identifies the bug at the calling statement rather than locating the actual source of the bug within the custom function itself. As shown in Fig.~\ref{fig:complicated_code}, this example exhibits a complex calling structure. The statements (\textit{i.e.}, colored in blue) call the custom ``combi'' and ``calc'' functions sequentially. Due to the complicated code structure, our Bug Locator only identifies the surface-level call statements, simply assuming that the bugs lie in the function parameters or subsequent value processing, without delving into whether there are logical bugs within the called functions themselves (\textit{i.e.}, colored in purple). The superficial localization that fails to address the root cause of the bug ultimately leads to a subsequent repair process that is also superficial, resulting in sub-optimal repair outcomes.

\subsection{RQ4. Human Study for Bug Localization.} 
\subsubsection{Qualitative Analysis of Bug Localization.} 
\begin{figure*}[t]
    \centering
    \includegraphics[width=\textwidth]{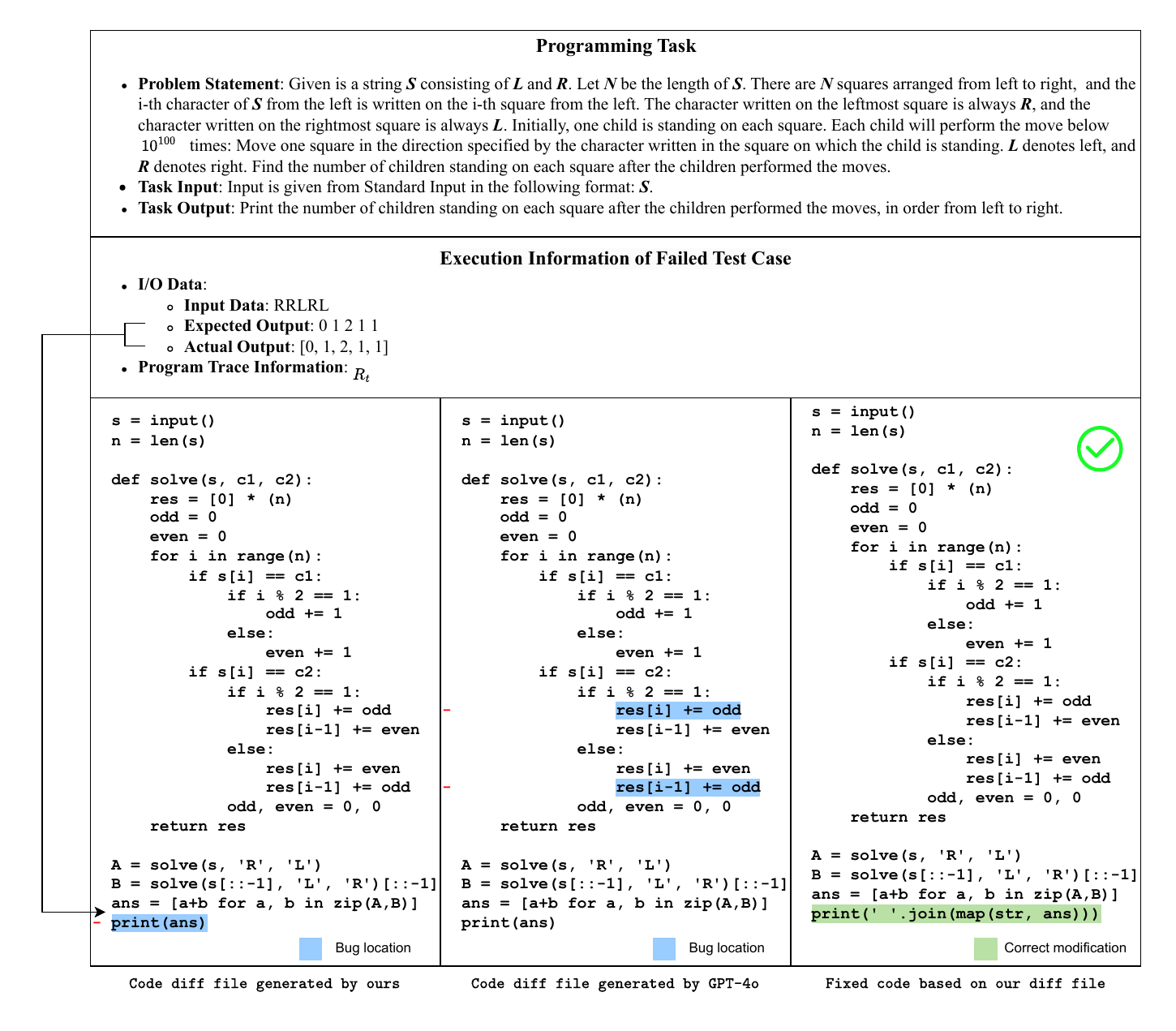}
    \caption{The comparison of accurate localization capabilities between GPT-4o and Ours.}
    \label{fig:qa-of-fl-1}
\end{figure*}

In the first stage, GPT-4o and CodeLlama trained under Self-Debug Learning generate \texttt{Code Diff} files. We conduct a manual evaluation to determine whether our method is helpful in locating bugs. The result shows that the CodeLlama trained under our framework is superior to the GPT-4o, including more accurate localization capabilities and more accurate semantic understanding.
(1) \textbf{Accurate Localization Capabilities}: 
As shown in Fig.~\ref{fig:qa-of-fl-1}, CodeLlama trained under our proposed method generates more accurate \texttt{Code Diff} files compared to GPT-4o. 
In the buggy code, the root cause of the bug is that the code handles the output format incorrectly. 
GPT-4o incorrectly pinpointed the issue to the statements ``res[i] += odd'' and ``res[i-1] += odd'', believing the bug is due to the incorrect calculation of intermediate values. 
In contrast, CodeLlama trained under our framework, accurately identifies that the problem lies with the final output statement. The superior accuracy is attributed to our framework, which trains the model to precisely locate bugs based on the differences between the expected output and the actual output. Even though GPT-4o has demonstrated ‌state-of-the-art performance in code-related tasks, it still lacks the ability to locate bugs accurately.
(2) \textbf{Semantically Accurate Understanding}: The second example shown in Fig.~\ref{fig:qa-of-fl-2}, which demonstrates CodeLlama under the proposed framework tends to generate more semantically accurate bug locations. The issue in the buggy code lies in logically associating the handling statements with the wrong branch conditions. 
Both \texttt{Code Diff} files seem correct at first glance, but a closer examination reveals that the one provided by GPT-4o is not semantically accurate, which attempts to change the $l$ that represents the left weight and the $r$ that represents the right weight, even though following its suggested \texttt{Code Diff} file may solve the problem. 
In contrast, CodeLlama trained under our framework directly marks out the handling statements in different branches as errors which is closer to the logic and semantics. 
Our framework enhances the model in considering semantic information more effectively when locating bugs.

\subsection{RQ5. Observed Experimental Phenomenon.} 
\subsubsection{Analysis of Ablation Study.} 
(1) Self-debug learning (SDL) is designed to identify runtime errors, mainly affecting the ACC/Improve metrics. SDL is introduced in the first stage to provide bug localization to support the second-stage repair. When GPT-4o is used in the second stage, the incorporation of SDL leads to a significant improvement in ACC (\textit{i.e.}, from 65.71\% to 67.57\%). However, the improvement is less pronounced when using CodeLlama in the second-stage. This is because the bug localization provided by SDL requires a powerful LLM (\textit{e.g.}, GPT4o) to accurately interpret and act accordingly, which explains the smaller performance gains of adding SDL with CodeLlama. 
(2) Regarding Adaptive Preference Learning (APL), we conducted an additional evaluation on larger code samples (where code exceeding 20 lines). Removing APL resulted in a significant decline in the Consistency metric (\textit{i.e.}, from 46.32\% to 44.43\%). This result highlights the effectiveness of APL in enhancing repair consistency, particularly in repairing longer code fragments. Given that developers often encounter complex tasks, APL is essential to boost model performance in such real-world scenarios.
\subsubsection{Analysis of model performance differences.} 
We discover that models with large parameters (\textit{e.g.}, gpt-4o, Claude-3.5) tend to rewrite the buggy code, deviating substantially from the original, which can explain why the consistency is low. 
Additionally, models with fewer parameters (\textit{e.g.}, CodeLlama-instrcut-7B) tend to merely copy the code without performing any edits.
\end{document}